# Fabrication of Hierarchical Sapphire Nanostructures using Ultrafast Laser Induced Morphology Change

Joshua Cheung, Kun-Chieh Chien, Peter Sokalski, Li Shi, Chih-Hao Chang[*]

*Walker Department of Mechanical Engineering, The University of Texas at Austin, Austin, TX 78712, USA*

[*]chichang@utexas.edu

## 1. Abstract

Sapphire is an attractive material in photonic, optoelectronic, and transparent ceramic applications that stand to benefit from surface functionalization effects stemming from micro/nanostructures. Here we investigate the use of ultrafast lasers for fabricating nanostructures in sapphire by exploring the relationship between irradiation parameters, morphology change, and selective etching. In this approach an ultrafast laser pulse is focused on the sapphire substrate to change the crystalline morphology to amorphous or polycrystalline, which is characterized by examining different vibrational modes using Raman spectroscopy. The irradiated regions are then removed using a subsequent wet etch in hydrofluoric acid. Laser confocal measurements conducted before and after the etching process quantify the degree of selective etching. The results indicate that a threshold laser pulse intensity is required for selective etching to occur. This process can be used to fabricate hierarchical sapphire nanostructures over large areas with enhanced hydrophobicity, which exhibits an apparent contact angle of 140 degrees and a high roll-off angle that are characteristic of the rose petal effect. Additionally, the fabricated structures have high broadband diffuse transmittance of up to 81.8% with low loss, which can find applications in optical diffusers. Our findings provide new insights into the interplay between the light-matter interactions, where Raman shifts associated with different vibrational modes can be used as a predictive measure of selective etching. These results advance the development of sapphire nanostructure fabrication, which can find applications in infrared optics, protective windows, and consumer electronics.

## 2. Introduction

Sapphire exhibits attractive properties including high mechanical hardness, broad optical transparency spectrum, and excellent chemical inertness.[1-3] These attributes make the material an ideal choice for photonic,[4-5] optoelectronic,[6-8] and functional transparent ceramic applications.[9-10] Many of these uses stand to benefit from surface functionalization using periodic arrays of micro or nanostructures. These structures can exhibit a variety of wetting states, including the Wenzel state, where the liquid fully wets the surface roughness and the Cassie-Baxter state,[11-12] where air pockets are trapped between the liquid and the structure. Both wetting states can be observed in hierarchical structures, known as the rose petal effect, where the liquid fully wets the microscale but not the nanoscale roughness.[13] Depending on the induced wetting state, these periodic arrays of nanostructures can exhibit superhydrophobic,[14-17] self-cleaning,[14,18-19] anti-fogging,[20-22] antibacterial,[23] or water harvesting properties.[23-25] Surface functionalization has been extensively demonstrated in conventional optics materials such as silica-based glass, which is well understood with established industry infrastructure. Unfortunately, the higher hardness and chemical stability of sapphire, which makes it desirable as a material, render many of the conventional micromachining processes ineffective.[26]

Currently, several techniques exist for patterning nanostructures in sapphire, including using a multilayer mask with etching,[27-28] nanomachining with an atomic force microscopy (AFM) probe,[29-30] or ablation using an ultrafast laser.[31-32] Multilayer masks employ silicon-based layers paired with reactive ion etching processes to increase etch selectivity and fabricate high aspect ratio periodic nanostructures. However, this approach requires complex deposition and etch steps to create and is limited to a small set of geometries and aspect ratios.[27-28] AFM nanomachining of sapphire utilizes a diamond probe to physically scribe into the surface of the substrate.[29-30] While this process allows for high-resolution patterning, the restrictions of the tip geometry and the low throughput of the serial process severely constrain practical applications. Ultrafast lasers, where high pulse intensities result in multiphoton absorption and material removal,[31-32] can be used to direct-write micro and nanostructures into sapphire substrates.[33-34] This technique allows for the micromachining of arbitrary geometries, even the creation of nanostructures below the substrate surface.[33-34] However the process suffers from defects and poor surface quality in certain conditions due to thermal and mechanical damage. In addition, the performance of ultrafast laser nanostructure fabrication depends heavily on the combination of ablation and morphology change mechanisms taking place, which is not well understood.

Ablation using ultrafast lasers with ns or ps pulses can induce high laser pulse intensities, on the order of $10^{13}$ W/cm$^2$,[35] with which the substrate is directly converted from a solid to a gaseous state. The ablation mechanism benefits from being a single step process with few pre-processing or post-processing requirements.[26] However, this process is often violent, with debris, cracks, and other defects resulting from the phase change and high thermal load inherent to ablation.[36-37] The ablation of sapphire via ultrafast laser has been thoroughly investigated in other works and damage thresholds with respect to pulse count and laser spectral characteristics can be experimentally identified.[38-41] Ultrafast laser irradiation has also been demonstrated to change the morphology of the sapphire substrate from single crystal to polycrystalline and amorphous.[42] The increase in grain boundaries in the polycrystalline and amorphous regions corresponds to higher etch rates with respect to the bulk substrate, with selectivity ratios as high as 1:10$^4$ being reported.[43-45] This difference in etch rates according to morphology allows for the maskless selective etching of the modified regions via dry or wet etching processes, resulting in the creation of surface structures.[46] Additionally, the lower thermal loads present and surface refinement inherent to the etching process results in fewer defects compared to ablation.[46] While other works have studied sapphire morphology change using transmission electron microscopy,[42] little research has been dedicated to understanding the effect of ultrafast laser parameters on the induced morphology change in sapphire. In addition, a knowledge gap exists between the morphology state of sapphire and its effect on selective etching.

In this work, we investigate the ultrafast laser induced morphology change of single-crystal sapphire using Raman spectrometry and harness this process to pattern hierarchical nanostructures. In this approach the micro-Raman measurements are employed to give insight into the crystallinity of the laser modified regions and identify the key vibrational modes that correspond to morphology change. Additionally, the irradiated sapphire regions are etched in hydrofluoric acid (HF) to better understand the relationship between morphology state and the degree of selective etching. Laser confocal measurements taken before and after the etch process allows the selective etching of the modified regions to be quantified with respect to laser irradiation conditions, the Raman peaks, and the morphology state. Lastly, the process is used to demonstrate the large-area surface functionalization of sapphire using ultrafast laser induced nanostructures. The resulting nanopatterned surface exhibits high contact angles and high roll-off angles, characteristics of rose petal effect, which is useful for antibacterial and water harvesting purposes. Additionally, specular and diffuse UV-Vis-NIR transmission measurements of the nanostructures demonstrates an efficient broadband optical diffuser. This work improves the understanding of sapphire morphology change via ultrafast laser and the corresponding selective etching, which can enable new sapphire nanomanufacturing processes with applications in photonics, optoelectronics, and transparent ceramics.

## 3. Experimental methodology

The ultrafast morphology modification and patterning process is illustrated in Figure 1. First, the single-crystal sapphire substrate in Fig. 1(a) is irradiated with ultrafast laser, resulting in the ablation and morphology change seen in Fig. 1(b). The irradiated region is then etched in HF to remove the polycrystalline areas, resulting in the structure shown in Fig. 1(c). The experimental setup used for the direct-write experiments is described in Supporting Information Section A.

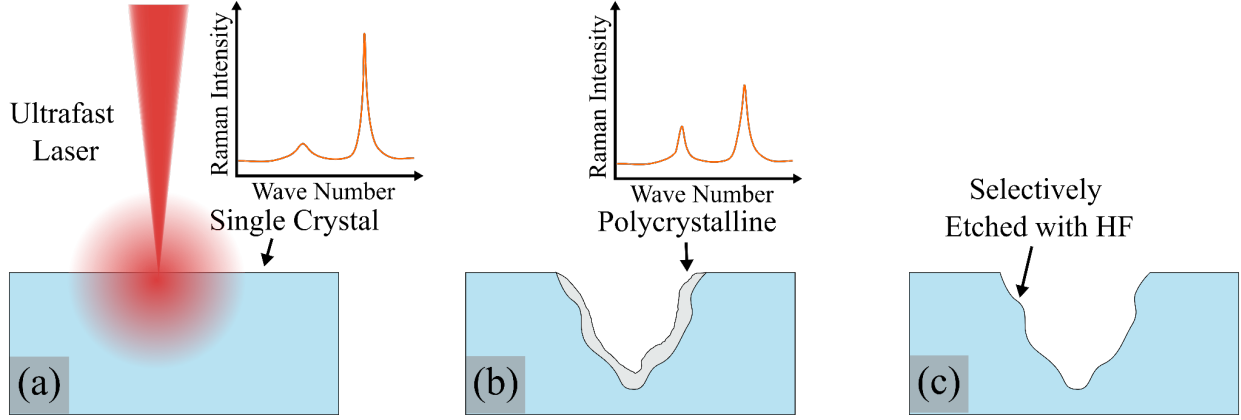

**Fig. 1: Proposed nanostructure fabrication process.** Diagram of (a) sapphire substrate irradiated with an ultrafast laser, (b) the resulting region with ablation and morphology change to polycrystalline, and (c) the structure after HF etching.

The effects of laser intensity and pulse count on morphology change and the subsequent selective etching in sapphire substrate is explored using a matrix of direct-write spots. Laser pulse count varies across the *x*-axis of the test matrix while laser pulse intensity varies across the *y*-axis of the test matrix. An adjustable-speed electronic shutter provides control over the pulse count. The pulse peak intensity $I_{peak}$ is determined by:

$$I_{peak} = \frac{P}{\tau \cdot A \cdot f} \tag{1}$$

where the average power *P* is measured using an optical power meter, spot area *A* is calculated based on the full width at half maximum (FWHM) beam diameter measured using the camera. Pulse duration $\tau$ and repetition rate *f* are 50 fs and 1 kHz, respectively, for the ultrafast laser used. The pulse intensity is controlled by altering the nominal power of the ultrafast laser, with all other parameters kept constant. Initially, a 5 x 5 reference grid is patterned onto the sapphire substrate using ablation to track the position of each irradiation spot. After creating the grid, direct-write spots are irradiated in the center of each of the spaces as illustrated in Fig. 1(a). The direct-write spots are patterned with laser parameters in accordance to their position in the test matrix. The laser pulse count varies linearly from 10 pulses to 50 pulses across the x-axis while the laser intensity varies exponentially from 160 TW/cm² to 2,560 TW/cm² across the y-axis.

To understand the morphology changes induced by the ultra-fast laser and the subsequent selective etching, the crystallinity and feature profiles, as shown in Fig. 1(b), are measured prior to the etching processes. The morphology is characterized by measuring the Raman spectra using a custom-built system alongside a commercial system with spatial mapping (Witec, micro-Raman Spectrometer Alpha 300). The Raman measurements are performed using a green laser with an excitation wavelength of 532 nm and an approximately 2 µm beam spot diameter. Following the pre-etch measurements, the sapphire wafer is immersed in 49% HF acid for 60 minutes at room temperature to selectively etch the regions with amorphous or polycrystalline morphologies, resulting in the structure shown in Fig. 1(c). The three-dimensional feature profiles are measured using a laser confocal microscopy system (Keyence, VK-X1100) before and after the HF etch to demonstrate selective etching. To quantify the degree of selective etching, the metric of etch area ratio $A_{ratio}$ was chosen, defined by:

$$A_{ratio} = \frac{A_f - A_i}{A_i} \qquad (2)$$

Where $A_i$ is the pre-etch surface cross-sectional area and $A_f$ is the post-etch surface cross-sectional area of the sample spot, as measured by confocal microscopy. An example measurement is presented in Supporting Information Section B.

**Macroscale surface functionalization of sapphire substrate**

An approximately 3.5 mm x 3.5 mm array of sapphire structures with a period of 7 µm were created to demonstrate hydrophobic effects resulting from surface functionalization. The ultrafast laser is scanned over a 2D dense grid with overlapping lines spaced 7 µm apart at a laser intensity of 1,155 TW/cm² and an average pulse count of 9.33 along the centerline. Following the ultrafast laser modification of the sapphire surface, the sample is etched in 49% HF acid for 60 minutes. The samples are then cleaned using RCA clean 1 (1:1:5 mixture of 29 wt% $NH_3$, 30 wt% $H_2O_2$, and $H_2O$) and oxygen plasma etching (18 W for 10 minutes) to prepare the sample for silane coating. The cleaned sample was placed inside a vacuum chamber along with a petri dish containing several drops of trichloro(octyl)silane (97%, Sigma Aldrich) for 8 hours, resulting in silane coated sapphire nanostructures. Following the application of a silane coating to reduce surface energy of the sapphire surfaces, contact angle measurements of the nanostructured and planar sapphire substrates were made using a goniometer (First Ten Angstroms, FTA 200). Optical properties were characterized with optical transmission measurements using a UV-Vis-NIR Spectrometer (Agilent, Cary 5000) with an external diffuse reflectance accessory using an integrating sphere (Agilent, DRA-1800).

## 4. Results and discussion

The matrix of irradiated ultrafast laser spots with various pulse intensity and count before and after HF etch are shown in the top-view SEM images shown in Fig. 2. Holes resulting from ablation can be observed in regions with a pulse intensity of 640 TW/cm$^2$ or higher as shown in Fig. 2(a). Furthermore, increasing irradiation intensity and pulse count both correspond to larger holes. Following HF etching, smoothing of the planar surface and some widening of the holes can be observed in Fig. 2(b). Additionally, cracks can be identified in the corners of the post-etch grid. Changes in the feature profiles and etched areas will be discussed in a later section.

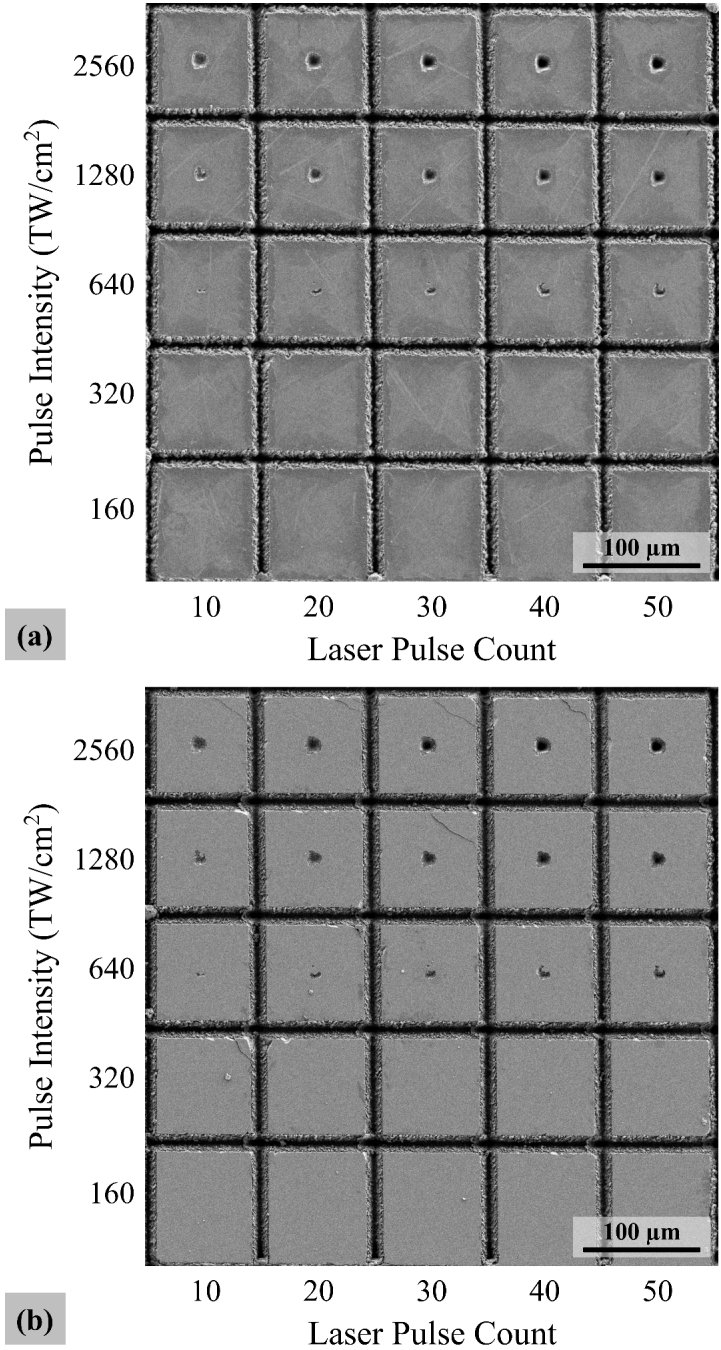

**Fig. 2: Matrix of samples before and after wet etching via HF acid.** (a) SEM image of the sample matrix before etching, with laser pulse count and laser intensity varying along the *x* and *y* axes, respectively. (b) Corresponding SEM image of the sample matrix after etching in HF for 60 minutes.

Morphology change of the crystalline sapphire to amorphous and polycrystalline states via ultrafast laser is a key mechanism to selective etching. The measured Raman spectra for the irradiated and pristine sapphire are shown in Fig. 3. For the sapphire substrate, the measured spectrum shows characteristic peaks at 382, 420, 433, 452, and 580 cm$^{-1}$. Notably, the 420 cm$^{-1}$ peak corresponds to the $A_{1g}$ vibrational mode while the remaining peaks exhibit an $E_g$ vibrational mode.[47-48] It can be observed that the pristine sapphire has a higher 420 cm$^{-1}$ peak and a lower 382 cm$^{-1}$ peak compared with the irradiated regions. It is important to note that due to variations in topography and the resulting differences in the scattering of the excitation light, direct comparisons of Raman intensities cannot be made accurately. Instead, the ratio of $E_g$ to $A_{1g}$ vibrational modes corresponding to 382 cm$^{-1}$ and 420 cm$^{-1}$ peaks will serve as the metric for comparing the degree of crystallinity for the irradiated regions to mitigate the effects of the aforementioned scattering. The 382 cm$^{-1}$ and 420 cm$^{-1}$ Raman peaks for the crystalline substrate have intensity counts of 61 and 1901 respectively, resulting in an $E_g$ to $A_{1g}$ ratio of 0.03. In contrast, for the most intense irradiated region with 2560 TW/cm$^2$ and 50 pulses, the 382 cm$^{-1}$ and 420 cm$^{-1}$ peaks have intensity counts of 101 and 465 respectively, resulting in a ratio of 0.22. This difference can be attributed to the selective excitation of different vibrational modes resulting from changes to the relative laser polarization direction with respect to the crystal.[48] Since the Raman measurement parameters are held constant, the change in the peaks corresponding to $E_g$ vs $A_{1g}$ modes indicates that the crystal orientation has been altered by the fs laser irradiation and is no longer c-plane sapphire, pointing to the polycrystalline nature of the irradiated spot. As such, lower $E_g$ to $A_{1g}$ ratios correspond to higher degrees of crystallinity, which is observed in the pristine sapphire. Understanding the relationship between the vibrational modes and morphology change helps to understand the parameters of laser pulse count and laser intensity, which are system specific, into morphology states which may be broadly applied.

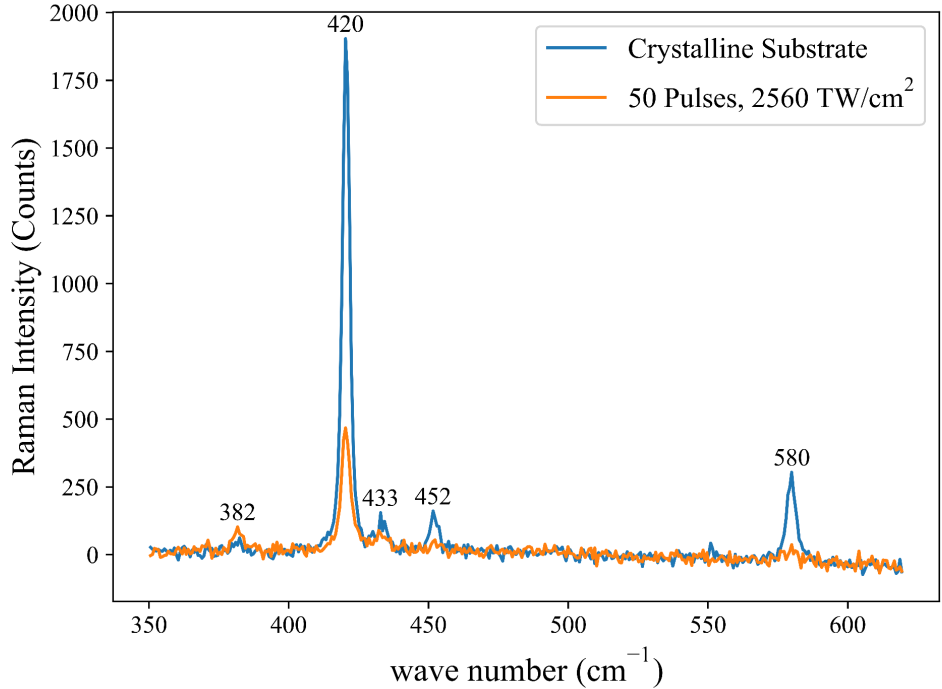

**Fig. 3: Comparison of Raman spectra between the crystalline substrate and an amorphous region.** The Raman spectra of the crystalline substrate and region irradiated with 50 pulses at 2560 TW/cm$^2$. Several distinct peaks for the c-plane sapphire substrate can be observed, with the most prominent being the 420 cm$^{-1}$ peak.

The measured Raman spectra for regions irradiated with varying intensity and pulse count are shown in Fig. 4(a) and Fig. 4(b), respectively. The pulse count is held constant at 20 for Fig. 4(a), where irradiation at low pulse intensities has little influence on the characteristic $E_g$ to $A_{1g}$ ratio as noted by the similar 382 cm$^{-1}$ and 420 cm$^{-1}$ peaks. A threshold intensity of around 640 TW/cm$^2$ can be noted before significant change in the peaks can be observed. Past the threshold, increases in pulse intensity further increases the 382 cm$^{-1}$ peak and decreases the 420 cm$^{-1}$ peak. The Raman spectra for spots with constant intensity of 1280 TW/cm$^2$ and variations in pulse counts are displayed in Fig. 4(b), from which the pulse counts appear to have little effect on either 382 cm$^{-1}$ or 420 cm$^{-1}$ peaks. The relatively high pulse counts present may have resulted in the saturation of morphology change effects, preventing any pulse count trends from being observed. The measured $E_g$ to $A_{1g}$ ratios for the entire experimental matrix are plotted in Fig. 4(c) to illustrate the degree of crystallinity of each irradiated spot. Based on the bar graph, the aforementioned threshold is observed across all pulse counts, reinforcing the existence of a threshold irradiation intensity required for morphology change.

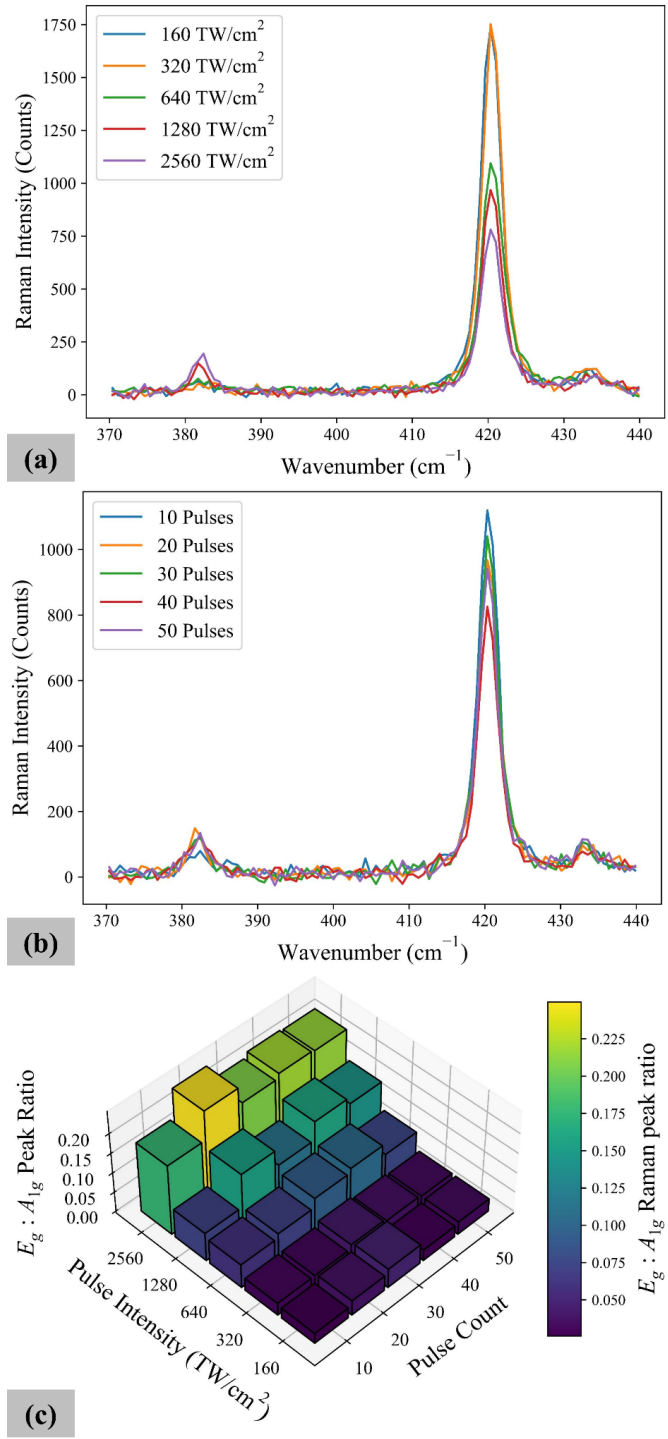

**Fig. 4: Micro-Raman measurements of the testing matrix.** (a) Measured Raman spectra for spots with 160 TW/cm$^2$ to 2560 TW/cm$^2$ peak intensity and constant 20 pulses. (b) Raman spectra or spots with 10

to 50 pulses at constant peak intensity of 1280 TW/cm$^2$. (c) Plot of the $E_g$ to $A_{1g}$ ratio for each irradiated region at the center of the spot.

In addition to the changes in morphology due to laser irradiation parameters, the spatial variation of the pattern spot is also examined, as shown in Fig. 5. Here the spot irradiated with 50 pulses at 1280 TW/cm$^2$ is used as a representative case. The changes in the measured Raman spectra moving from location 1, with higher laser irradiation, to location 5, which features a morphology similar to that of the crystalline substrate can be seen in Fig. 5(a). The decrease in the $A_{1g}$ mode can be observed in the 420 cm$^{-1}$ peak, while changes to the $E_g$ mode located at 382 cm$^{-1}$ are less discernible. The corresponding locations across the spot boundary are depicted in the SEM image shown in Fig. 5(b), which demonstrates the short length scales of approximately 5 µm in which the transition from amorphous and polycrystalline to fully crystalline occurs. The $E_g$ to $A_{1g}$ ratio of the five locations are plotted in Fig. 5(c), with a trend of increasing crystallinity moving from location 1 to location 5.

The spatial variation of morphology for the matrix of spots is investigated in Fig. 5(d), using micro-Raman mapping of the 420 cm$^{-1}$ peak intensities. Despite not being as robust of a metric as $E_g$ to $A_{1g}$ ratios, 420 cm$^{-1}$ peak intensities are used in the micro-Raman mapping due to the sensitivity limitations of the micro-Raman Spectrometer Alpha 300 in measuring the smaller 382 cm$^{-1}$ peaks. The map of each spot has been linearly interpolated with respect to crystalline sapphire and the most amorphous measurement, allowing for comparison between the spots. The lack of discernible spots at lower intensities supports the notion of a threshold irradiation intensity for morphology change to occur. Irradiated regions above the intensity threshold feature a mostly uniform amorphous or polycrystalline spot with a short transition to the crystalline background. This reinforces the short transition length scales found in Fig. 5(b).

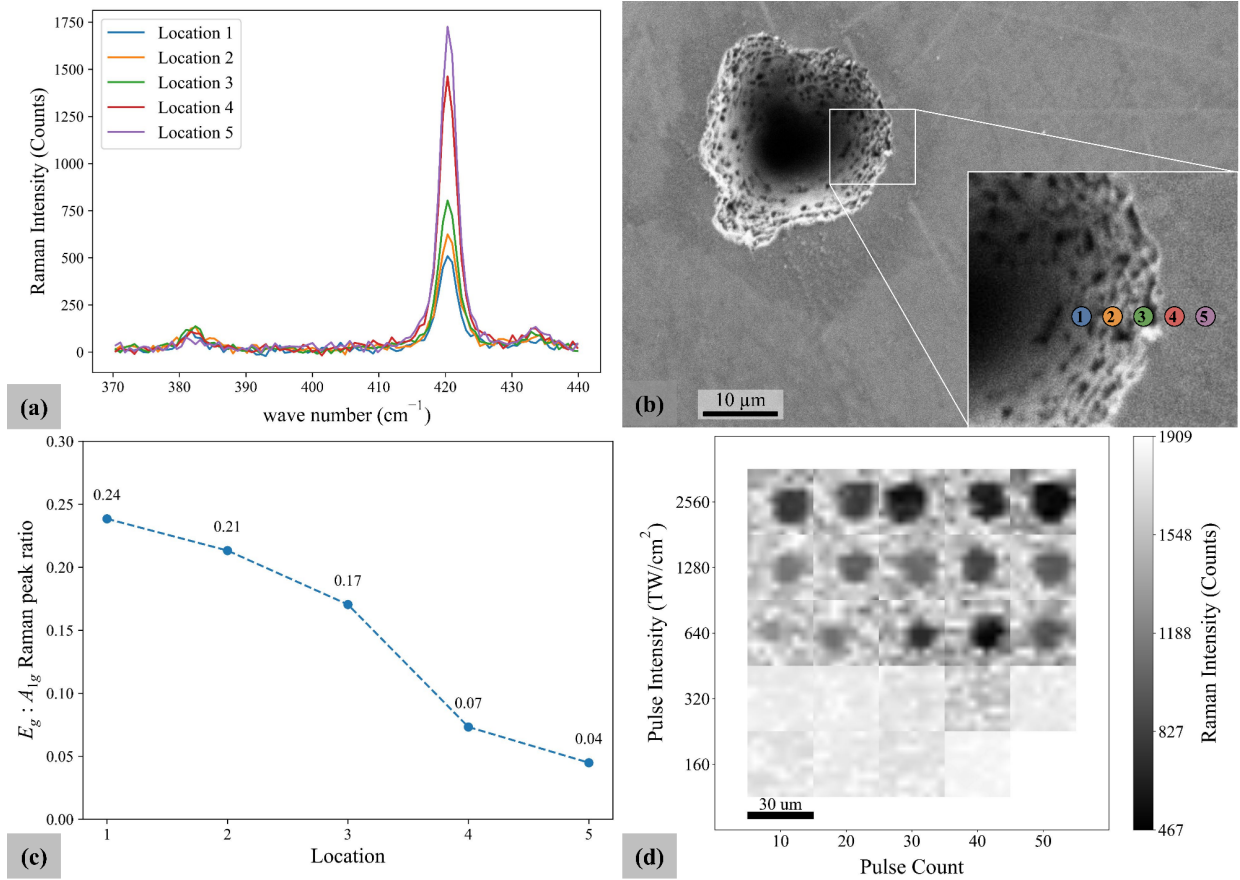

**Fig. 5: Micro-Raman measurements from along the transition from amorphous and polycrystalline to crystalline.** (a) Micro-Raman spectra for each of the five locations across the boundary for the 1280 TW/cm$^2$ pulse intensity and 50 pulse spot. (b) Pre-etch SEM image of the irradiated spot depicting the corresponding locations denoted in part a. (c) Plot of the $E_g$ to $A_{1g}$ ratio at each location. (d) Plot of the micro-Raman spatial mapping of the 420 cm$^{-1}$ peak for each of the spots. Each map is 30 μm wide, centered on the irradiated spot, the rest of the grid is not pictured.

With the effects of laser parameters and spatial variation on morphology better understood, the impacts on selective etching are investigated next. Fig. 6(a) and Fig. 6(b) below shows the 1280 TW/cm$^2$ and 50 pulses region before and after the HF etch. The etch selectivity of concentrated HF for crystalline vs polycrystalline/amorphous is as high as 1:10$^4$, ensuring only the material with modified morphology will be removed.[43] Note the improved surface finish and the removal of material at the edges of the reference grid as a result of the etch. The edges of the spot are similarly etched, and pores can be observed inside the spot by the selective removal of amorphous and polycrystalline material. Analysis of Fig. 2(a) and Fig. 2(b) reveals a similar trend across the entire matrix of spots, with effects of the HF etch observed in the widening of the reference grid lines and the smoothing of the plane surface.

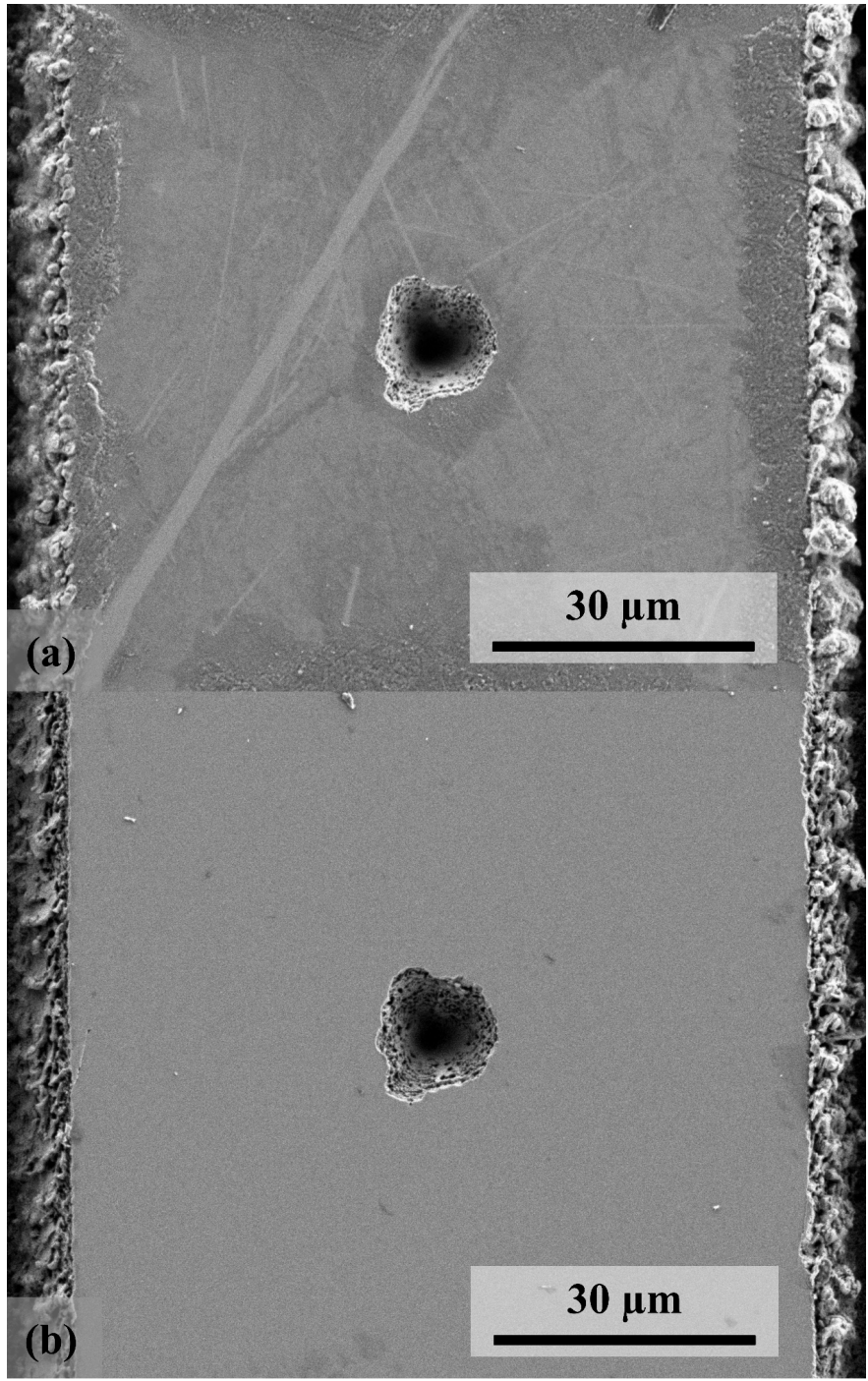

**Fig. 6: Top-View SEM of 1280 TW/cm² and 50 pulses irradiated region before and after etch.** (a) SEM image of the spot irradiated with 1280 TW/cm² and 50 pulses before etching. Note the marks and rough surface present. (b) SEM image of the spot post etch.

To better understand selective etching due to morphology change, the profile of the pattern can be extracted from the confocal microscope data, as shown in Figure 7. Here Fig. 7(a), 7(b), and 7(c) depict the pre-etch and post-etch cross-sectional profiles for the 640, 1280, and 2560 TW/cm$^2$ spots at 20 pulses, respectively. The post-etch profiles feature a slight broadening of the width when compared to the pre-etch profiles, with the width increasing by approximately 0.83, 1.1, and 1.4 µm for the 640, 1280, and 2560 TW/cm$^2$ spots, respectively. A similar increase in depth is observed, where the post-etch profile is deeper by approximately 0.4, 0.7, and 0.8 µm for the corresponding samples. The contours appear to maintain similar topographic profiles before and after the HF etch, likely a result of the isotropic nature of wet etch processes as well as the relatively small etch depths present. Note the high surface roughness present on the surface of the pre-etch profiles, a result of the defects and ejecta resulting from laser irradiation. In contrast, the smoother surface profiles for the post-etch spots are caused by the removal of the polycrystalline and amorphous material in the wet-etch process.

The etch area ratio for each spot in the testing matrix is shown in Fig. 7(d). Note the lack of etch area ratio for irradiated regions with laser intensities of 160 TW/cm$^2$ and 320 TW/cm$^2$, indicating a minimum pulse intensity threshold before any selective etching can occur. Other metrics were explored, such as the change in maximum depth of the spots before and after etching. However, difficulties were encountered in resolving the bottom of the high pulse count and intensity spots using laser confocal microscopy as demonstrated in Supporting Information Section C. Across the irradiated regions, the maximum measured depth was 14.1 µm, at which point the reflection signal was too low. Paired with the high noise levels inherent to point-based measurements, these two measurement limitations guided the choice of etch area ratio as the preferred metric for selective etching.

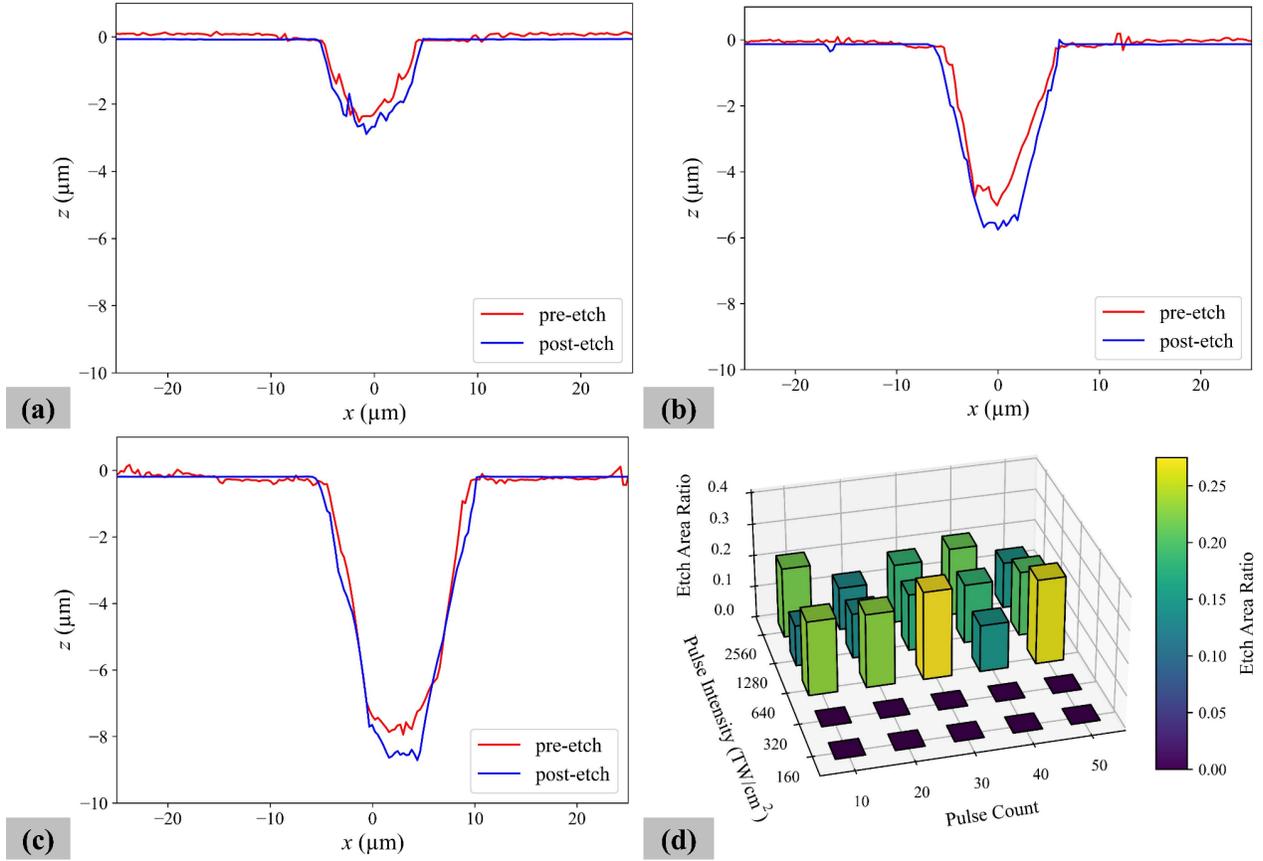

**Fig. 7: Selective etch results using confocal microscopy.** Measured pre-etch and post-etch profiles for the 20 pulses spots with irradiation intensity of (a) 640, (b) 1280, and (c) 2560 TW/cm$^2$. (d) Bar graph detailing the etch area ratio for each of the 25 combinations of pulse intensity and pulse count parameters.

The relationship between laser irradiation parameters and etch area ratio are reinforced in Fig. 8. Here the etched area ratios are plotted versus pulse intensity and pulse count in Fig. 8(a) and 8(b), respectively. The previously observed threshold intensity is supported in Fig. 8(a), where selective etching does not occur at pulse intensities below 640 TW/cm$^2$. In pulse intensity regimes where selective etching takes place, there is no significant correlation between pulse intensity and etch area ratio, with no recognizable patterns in the constant pulse count lines past the threshold, as shown in Fig. 8(a). Furthermore, pulse count does not have a significant influence on etch area ratio, as indicated by the lack of clear trends in the constant pulse intensity lines in Fig. 8(b). One explanation for this result can be the relatively high pulse count floor of 10 tested, where the morphology change effect has already saturated. The pulse number is expected to have an effect on the selective etch and morphology change at lower pulse counts below 10, which is the subject of on-going efforts. The etch area ratio is plotted versus the measured $E_g$ to $A_{1g}$ ratio in Fig. 8(c) and corroborates the existence of a morphology state threshold necessary for selective etching to occur. Moreover, a further increase in Raman $E_g$ to $A_{1g}$ ratios beyond the threshold appears to have less effect on etch area ratio. As such, Fig. 8(c) describes a binary relationship between selective etching and the morphology state, with two distinct etching and non-etching phases. This result also indicates that the crystallinity measured by Raman spectra can be an accurate predictor of the final features after wet etching. Additional examples of the observed trends can be seen in the SEM images, confocal height maps, and Raman spectra of the 20 pulse spots found in Supporting Information Section D.

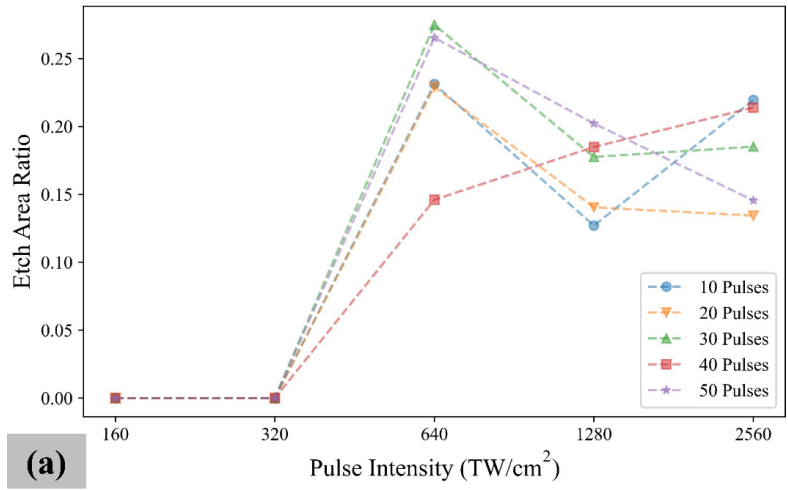
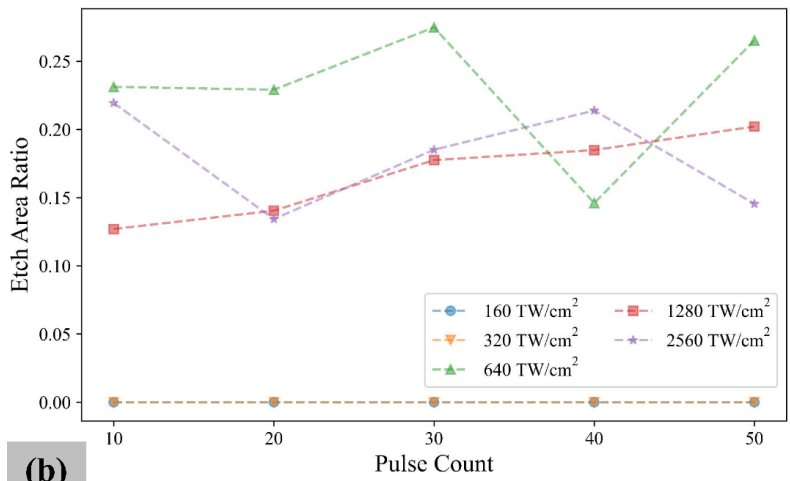
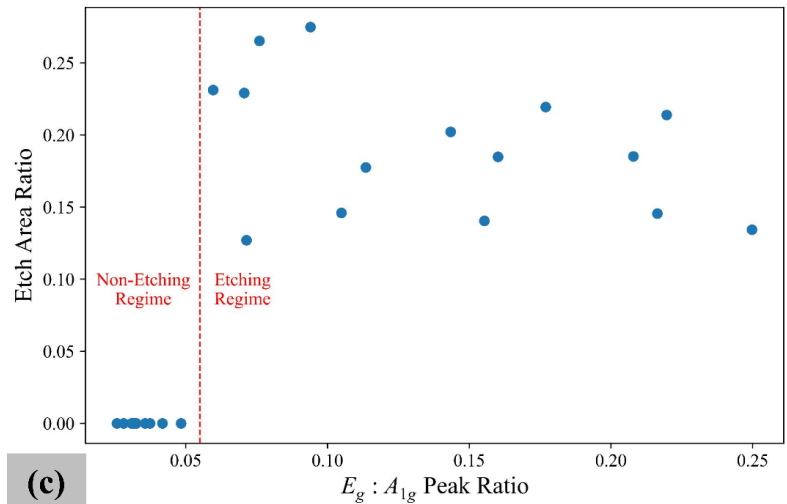

**Fig. 8: Correlations between laser parameters and Raman spectra to selective etching.** (a) Graph of etch area ratio as a function of (a) pulse intensity and (b) pulse count. (c) Graph of etch area ratio as a function of Raman $E_g$ to $A_{1g}$ ratio measured using Raman microscopy.

**Hydrophobic surface functionalization demonstration**

To demonstrate surface functionalization using ultrafast laser induced morphology change and the subsequent selective etching, sapphire nanostructures are patterned over an area of 12.25 mm$^2$, as seen in Fig. 9(a). The SEM image of the periodic nanostructures shown in Fig. 9(b) illustrates the hierarchical structure, with nanometer-scale roughness present on top of the underlying periodic microstructures. The period and height of the microstructure are measured to be 7 and 5.7 µm, respectively. The surface was measured with confocal microscopy to have a roughness factor of 2.34. However, the measurements are limited by the system resolution and do not account for the nanometer-scale features, as seen in the height map found in Supporting Information Section E. This surface roughness is a critical component of hydrophobicity in accordance with the Wenzel equation [15]

$$\cos \theta^* = r\cos \theta \quad (3)$$

where $\theta$ is the contact angle (CA) with a flat surface, $r$ is the roughness factor, and $\theta^*$ is the apparent contact angle. The contact angle for the silane coated flat sapphire substrate was measured in Fig. S10(a) to be 106 degrees, which is hydrophobic. The contact angle of the silane coated sapphire nanostructures was measured at 140 degrees in Fig. S10(b), exhibiting a marked improvement of 34 degrees over the sapphire substrate due to the hierarchical structure. Moreover, the nanostructures have a higher contact angle than the 130 degrees predicted by the Wenzel model. Such improvements in contact angle, nearing the superhydrophobic regime of 150 degrees, can be attributed to the hierarchical nature of the sapphire structures. Roll-off angle measurements of the hierarchical sapphire nanostructures in Fig. 9(c) indicate that the drop does not roll off even when perpendicular, indicating high contact angle hysteresis (CAH). The high liquid adhesion paired with the high static CA is characteristic of the rose petal effect, which can be attributed to the liquid wetting the microstructure but not the nanoscale roughness.[19,23-24] The unique combination of high CA and high CAH has applications in antibacterial, water harvesting, and guided fluid transport surfaces.[19,23-25] Detailed goniometer measurements are described further in Supporting Information Section F.

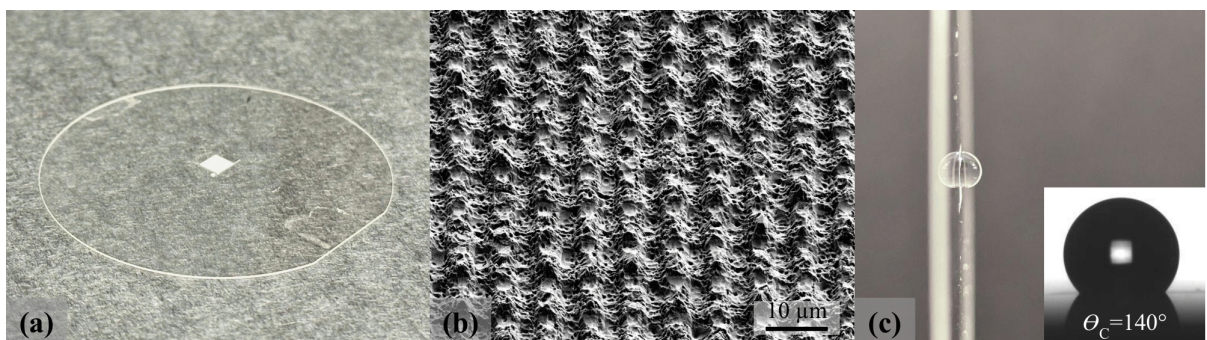

**Fig. 9: Demonstration of hydrophobic effects induced by sapphire nanostructures created via ultrafast laser.**
(a) Image of the 3.5 mm by 3.5 mm patch of sapphire nanostructures created using ultrafast laser irradiation (center of 100 mm wafer). Subsequent HF etching and silane coating were performed to induce selective etching and increase water contact angle respectively. (b) Angled-view SEM image of post-etch nanostructures with hierarchical features. (c) Demonstration of high contact angle hysteresis of the post-etch sapphire nanostructures coated in silane. An inset image shows the contact angle measurement for the sample.

The optical transmission characteristics of sapphire are also critical for many applications and are characterized for the fabricate sample. The specular and diffuse transmittance measurements for the planar and nanostructured sapphire substrates are illustrated in Fig. 10. The specular and diffuse transmittance of the sapphire substrate have an average of 83.2% and 4.1%, respectively, across the range of wavelengths from 250 nm to 2500 nm. The high specular and low diffuse transmittance for the sapphire substrate is characteristic of optical windows, as expected from the polished surfaces. Conversely, the nanostructures demonstrate the opposite behavior, with an average specular and diffuse transmission of 11.9% and 73.9%, respectively. A peak diffuse transmittance of 81.8% can be observed at 1354 nm wavelength and maintains this value until gradually falling at longer wavelengths. The micro/nanoscale roughness of the hierarchical structures induces light scattering that results in predominantly diffuse transmission. Note that the total transmission of the planar and nanostructured samples across the measured wavelength are nearly identical as described in Supporting Information Section G, with averages of 87.4% and 85.8%, respectively. The total transmission of the nanostructured sapphire sample is even higher than the substrate at wavelengths longer than 1600 nm, reaching a peak of 96.1% at 2135 nm. Consequently, the nanostructures exhibit 98.2% of the substrate total transmission, demonstrating low optical losses. Furthermore, the flat trends of the nanostructure transmittance measurements across the visible to infrared point to the broadband nature of their optical behavior. The efficient broadband diffuse characteristics of the sapphire nanostructures make them suitable for applications in laser illumination,[49] optical sensors,[50] and display technologies.[51]

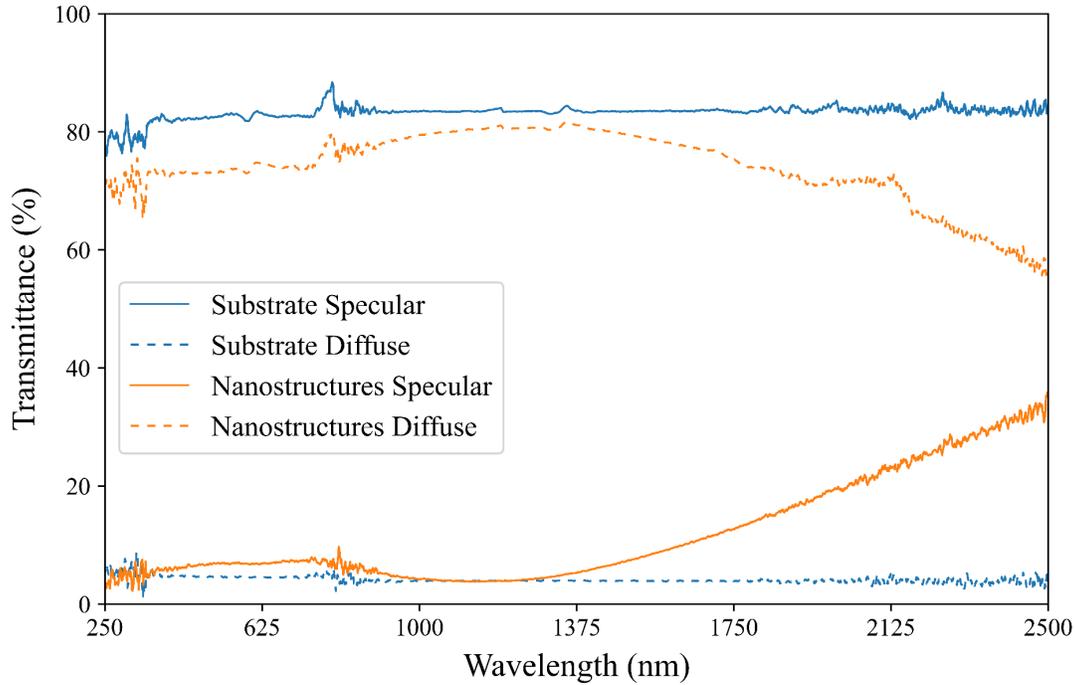

**Fig. 10: Optical transmittance properties of sapphire nanostructures created via ultrafast laser.** Graph of diffuse and specular measurements of the post-etch sapphire nanostructures and sapphire substrate.

## 5. Discussion and future work

The laser irradiation results demonstrated clear relationships between irradiation parameters and the subsequent selective etching in HF, namely the presence of a threshold pulse intensity around 640 TW/cm$^2$ necessary for etching to occur. The data also indicates the lack of any strong trends between pulse count and selective etching, which can be attributed to the relatively high pulse numbers tested. The degree of morphology change can also be precisely characterized by using Raman microscopy, where the ratios of the intensity peaks of the $E_g$ to $A_{1g}$ vibrational modes correlates highly to selective etching. It is important to note the increments in pulse intensities used in this study are too coarse to accurately identify the exact threshold value, and more experiments between 320 TW/cm$^2$ and 640 TW/cm$^2$ are needed. Likewise, further experiments at lower pulse count regimes, namely between 1-10 pulses, are needed to avoid saturation and effectively investigate light-matter interactions within a single pulse. Additional material characterization using transmission electron microscopy and electron beam diffraction can provide morphology measurements with greater spatial resolution. By continuing to explore the relationship between morphology state, irradiation parameters, and selective etching, ultrafast laser induced selective etching can be better understood to improve the fabrication processes.

Additionally, while the macroscale sapphire nanostructure sample successfully demonstrated increased hydrophobicity by adding hierarchical roughness, further experiments are required to better understand and control the surface geometry. One key question is whether the microscale structure and nanoscale roughness can be independently controlled, which will be examined by varying the beam overlap during irradiation. Furthermore, the same roughness responsible for the nanostructures' hydrophobic effects also results in high diffuse and low specular transmission, characteristics that are undesirable for windows but are useful for optical diffusers. This roughness is a product of the violent nature of the ultrafast laser, an issue compounded by the selectivity of HF, reported to be as high as $1:10^4$ between crystalline and amorphous sapphire.[43] This high selectivity leaves crystalline regions virtually unetched, resulting in sharp edges and pits where local morphology varies. To address this problem, alternative etch processes will be explored to reduce the nanoscale roughness, namely 78% $H_2SO_4$ + 22% $H_3PO_4$ at 270 °C, with a lower selectivity of approximately 1:66 and a higher etch rate.[43] By further exploring the mechanisms of hydrophobicity present in our sample and experimenting with etching processes, it is possible to improve the optical clarity of the fabricated sapphire nanostructures while maintaining their surface functionalization.

## 6. Conclusion

In this work, we investigate the relationships between ultrafast irradiation parameters and selective etching and demonstrate ultrafast laser fabrication of sapphire nanostructures with enhanced hydrophobicity and light scattering properties. Using Raman microscopy, the ratio of the $E_g$ to $A_{1g}$ vibrational modes can identify a threshold pulse intensity necessary for selective etching to take place. Additionally, a binary relationship between the morphology state of the irradiated regions and the degree of selective etching can be observed. The results indicate that the Raman spectra can be used to quantify the degradation in crystallinity and serve as a predictive metric for selective etching. We also successfully demonstrated the fabrication of sapphire nanostructures over macroscale areas via ultrafast laser and selective etching. The surface functionalization of the sample exhibited a high apparent contact angle of 140 degrees and high contact angle hysteresis, a demonstration of the rose petal effect. Furthermore, the hierarchical surface roughness accompanying the nanostructures resulted in efficient diffuse transmission of up to 81.8% and can have applications in broadband optical diffusers. Future works will focus on investigating irradiation parameters and morphology states near the threshold to better refine our understanding of selective etching using ultrafast lasers. Additionally, we will explore secondary etching processes to control the optical properties and hydrophobic effects of the nanostructures.

## 7. Acknowledgements

This work was performed at UT Austin Texas Materials Institute (TMI), the Nanomanufacturing System for mobile Computing and Energy Technologies (NASCENT), and Texas Nanofabrication Facilities, which is supported by the National Science Foundation (NSF) as part of the National Nanotechnology Coordinated Infrastructure (NNCI) grant NNCI-2025227. This work is funded by the Army Research Office (ARO) grant W911NF-22-1-0124 and the National Science Foundation (NSF) grant CBET-2343526.

# Supporting Information for "Fabrication of Hierarchical Sapphire Nanostructures using Ultrafast Laser Induced Morphology Change"

**Authors:** Joshua Cheung, Kun-Chieh Chien, Peter Sokalski, Li Shi, Chih-Hao Chang

**Affiliations:**

Walker Department of Mechanical Engineering, The University of Texas at Austin, Austin 78712

# Supporting Information A: Experimental Setup

The experimental setup used for the direct-write experiments is illustrated in Fig. S1. An ultrafast laser (Spectra-Physics Solstice Ace) has a nominal pulse duration of 50 fs, a repetition rate of 1 kHz, and a wavelength of 790 nm. A dichroic mirror, which reflects infrared light and transmits visible wavelengths, is used to direct the laser into the objective, while allowing the emitted light from the ablated spot to be imaged in-situ by the camera. The objective lens (Carl Zeiss, EC Plan-Neofluar 10x/0.3) is used to focus the ultrafast laser onto the sapphire wafer (MSE Supplies, 2 inch sapphire wafer c-plane, double side polish 300 um) that is positioned via a stage which can translate in the $x$, $y$, and $z$ axes.

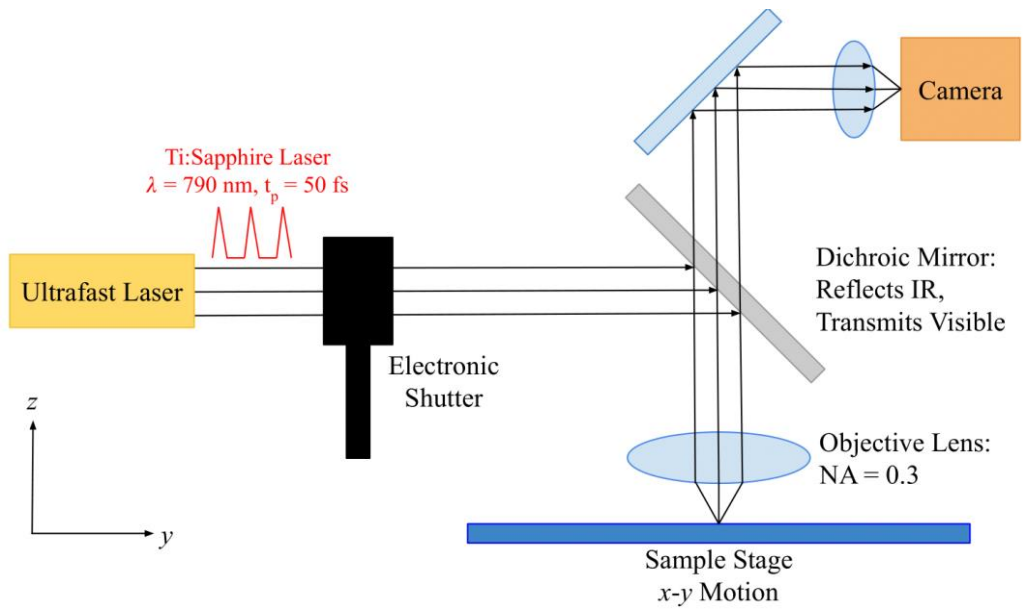

**Fig. S1: Direct-write experimental setup.** Used for creating a matrix of direct-write spots and a periodic array of structures.

# Supporting Information B: Example Measurement of Etch Area Ratio

The example measurement of etch area ratio for the 50 pulses 2560 TW/cm² irradiation spot is shown in figure S2. Laser confocal measurements before and after etching are shown in Fig. S2(a) and Fig. S2(b) respectively. For each measurement, a threshold height is selected to maximize the number of pixels in the irradiated spot without including the planar sapphire substrate. The number of red pixels is directly proportional to the surface area and is used to calculate an etch area ratio of 0.145 according to Eq. 2:

$$A_{ratio} = \frac{A_f - A_i}{A_i} = \frac{2787 - 2433}{2433} = 0.145$$

Fig. S2(c) demonstrates a case where the threshold height is set 12 nm above the selected threshold height in Fig. S2(b), a difference equivalent to the Keyence's stated measurement repeatability using the 50x objective. Note the appearance of many red dots on the planar surface away from the irradiation region signifies that the threshold height is too high and no longer selecting for the irradiated spot. The distinct transition from measuring pixels in the irradiated spot to including planar pixels is displayed in Fig. S2(d), where a sharp rise in red pixel counts can be seen right after the selected threshold height.

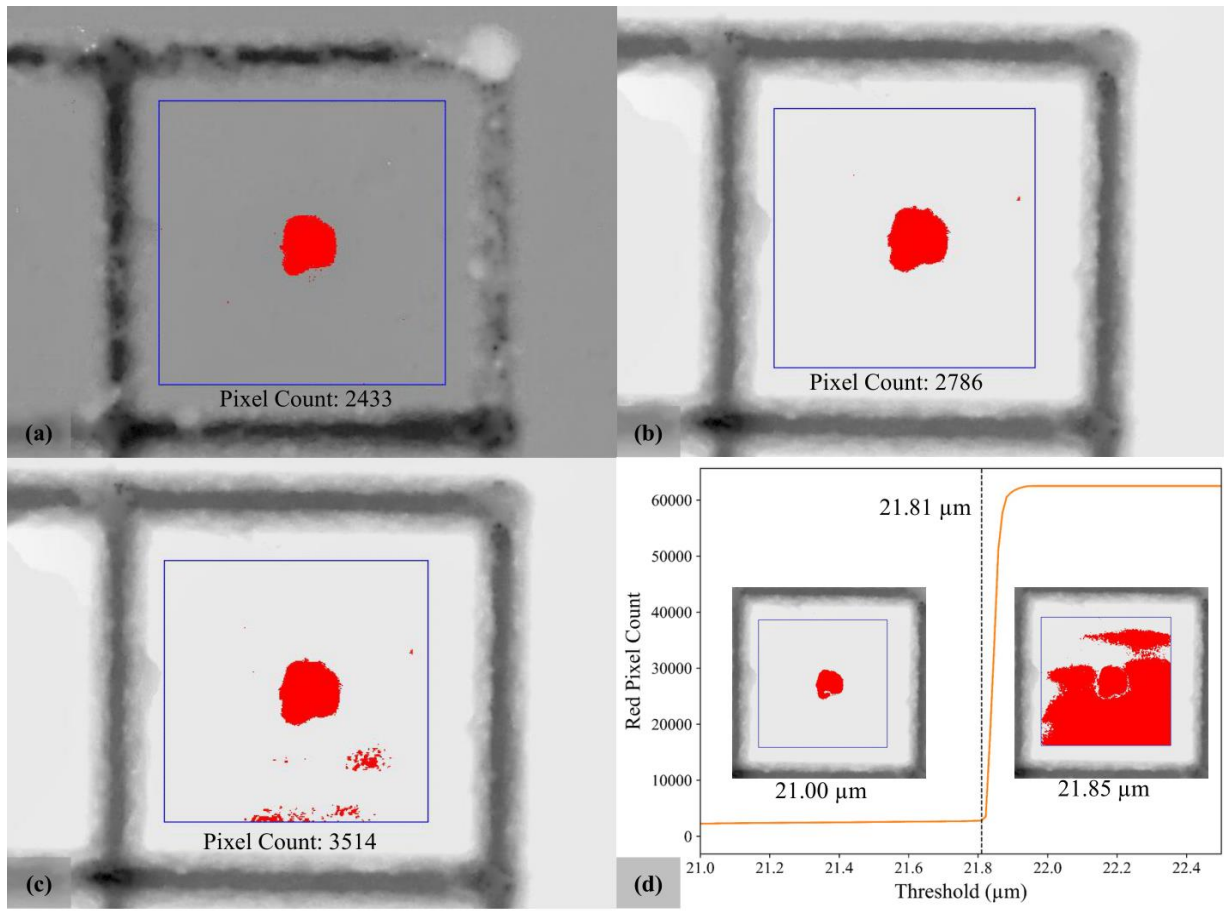

**Fig. S2: Selective etch results using confocal microscopy.** Laser confocal microscopy images of the 2560 TW/cm² and 50 pulses irradiation spot (a) before and (b) after etching with threshold heights of 16.06 µm and 21.81 µm respectively. The blue box is the bounding limits of the pixel measurement area, while the red pixels are those below the height threshold. (c) Image of the post-etch 2560 TW/cm² and 50 pulses irradiation spot with a threshold height of 21.822 µm. (d) Graph of red pixel count as a function of threshold height for the post-etch 2560 TW/cm² and 50 pulses irradiation spot, with inset images for thresholds above and below 21.81 µm.

# Supporting Information C: Example of Laser Confocal Measurement Issues for High Pulse Count and Intensity Samples

In the case of high pulse count and high intensity samples, difficulties were encountered in resolving the bottom of the irradiation regions using laser confocal microscopy. The 2560 TW/cm$^2$ and 50 pulses is a clear example of this issue, and is displayed in Fig. S3 below. A deep hole is indicated by the pre-etch and post-etch SEM images in Fig. S3(a) and Fig. S3(b) respectively. However, the laser confocal is unable to resolve the bottom of the hole as shown in the pre-etch and post-etch height maps in Fig. S3(c) and Fig. S3(d), respectively. This can be attributed to the increased depth of the hole and the resulting low reflected light signal.

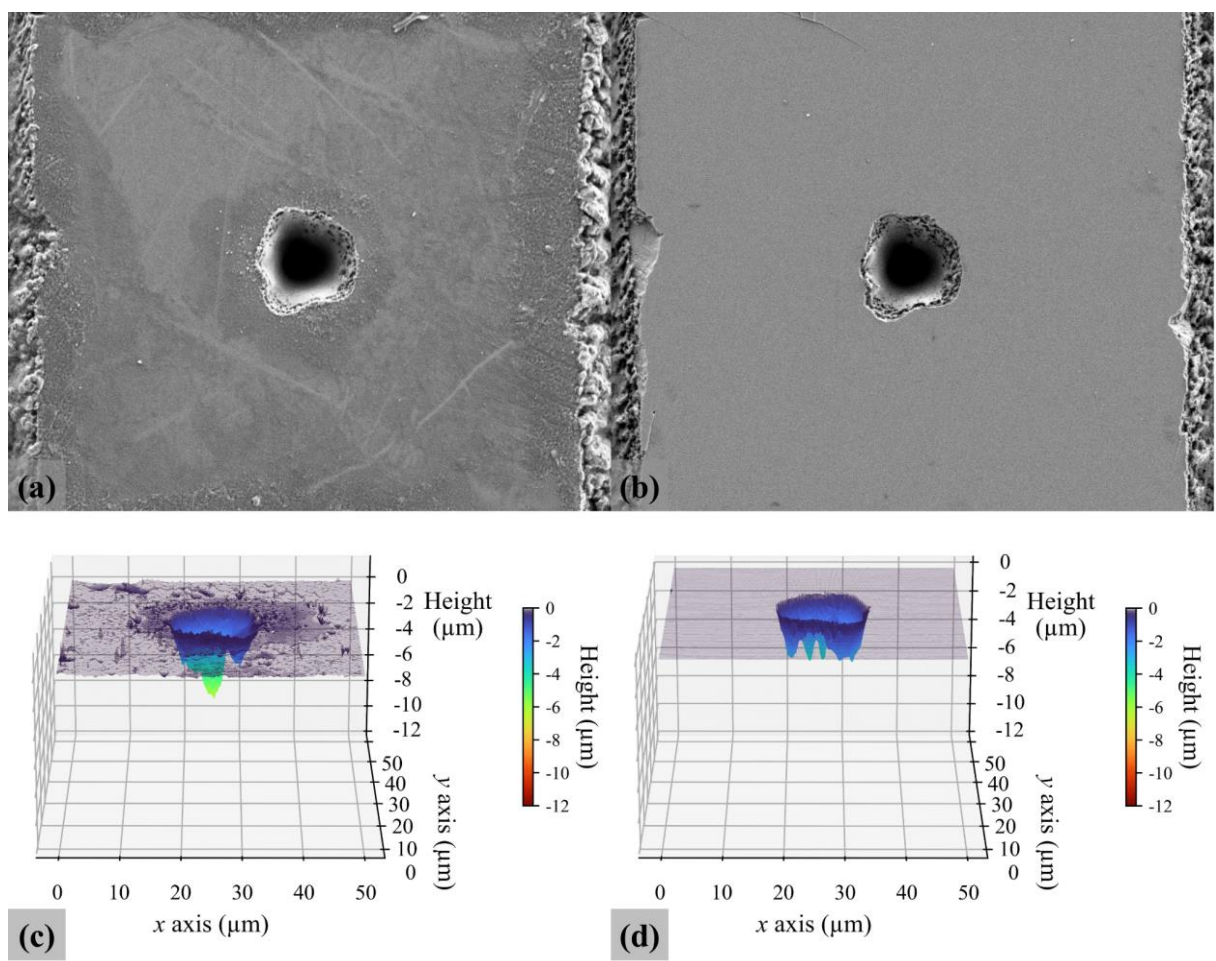

**Fig. S3: Example of laser confocal difficulties in the 2560 TW/cm$^2$ and 50 pulses sample.** SEM images of the (a) pre-etch and (b) post-etch irradiated region with the corresponding (c) pre-etch and (d) post-etch height maps created from laser confocal measurements.

# Supporting Information D: SEM Imaging and Laser Confocal Measurements of 20 Pulses Spots

The raw measured data for confocal, Raman, and SEM images for the 20 pulses irradiation spots for intensities of 160, 320, 640, 1280, and 2560 TW/cm$^2$ are shown in Fig. S4-S8, respectively. Each figure displays the pre-etch and post-etch SEM and height map image. A shared height map scale bar allows for the direct comparison of laser confocal measurements between samples. The micro-Raman spectra of the center of each irradiated spot is shown with key peaks highlighted.

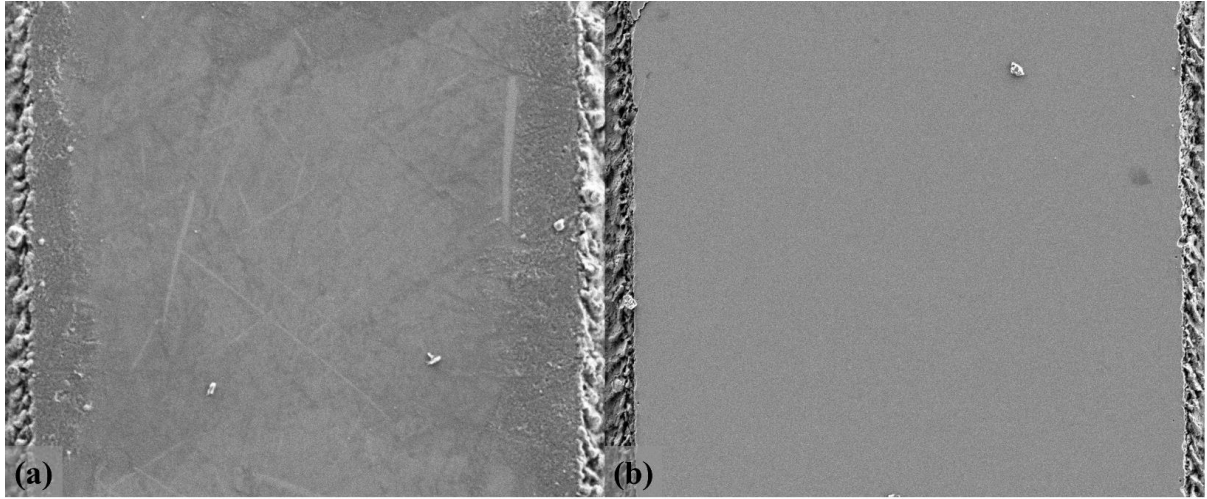
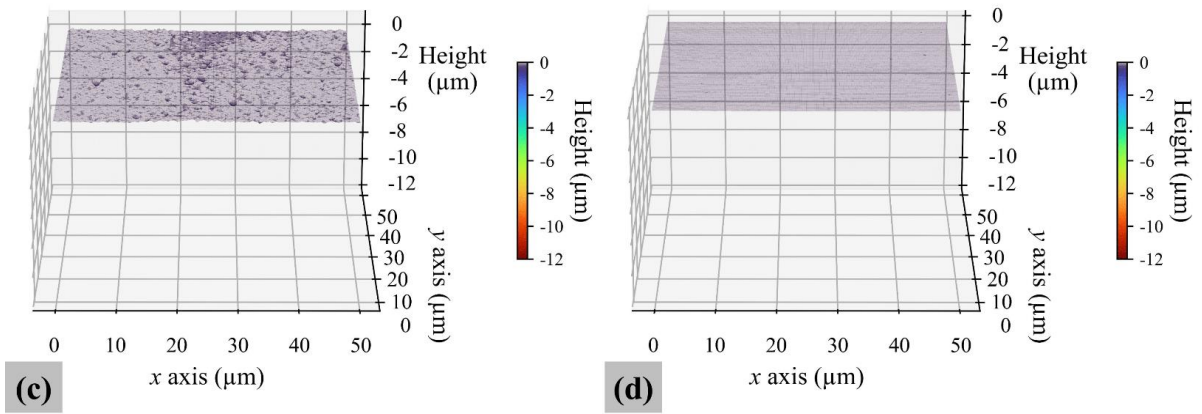
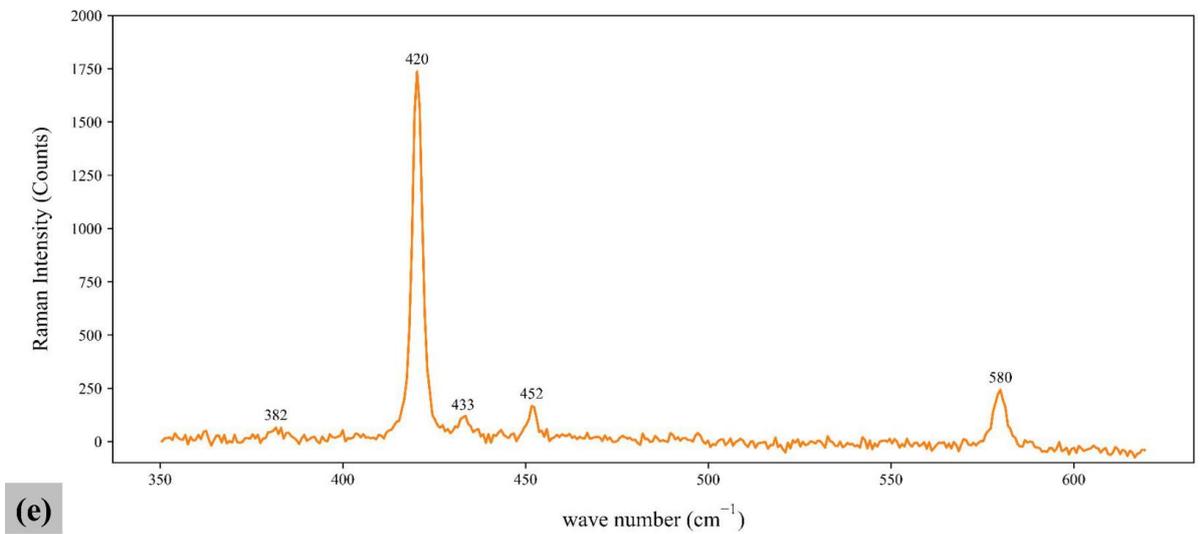

**Fig. S4: Metrology of 160 TW/cm$^2$ and 20 pulses sample.** The measurements of the 160 TW/cm$^2$ and 20 pulses spot include (a) pre-etch and (b) post-etch SEM images, (c) pre-etch and (d) post-etch laser confocal height maps, and (e) pre-etch micro-Raman spectra for the center of the irradiated region.

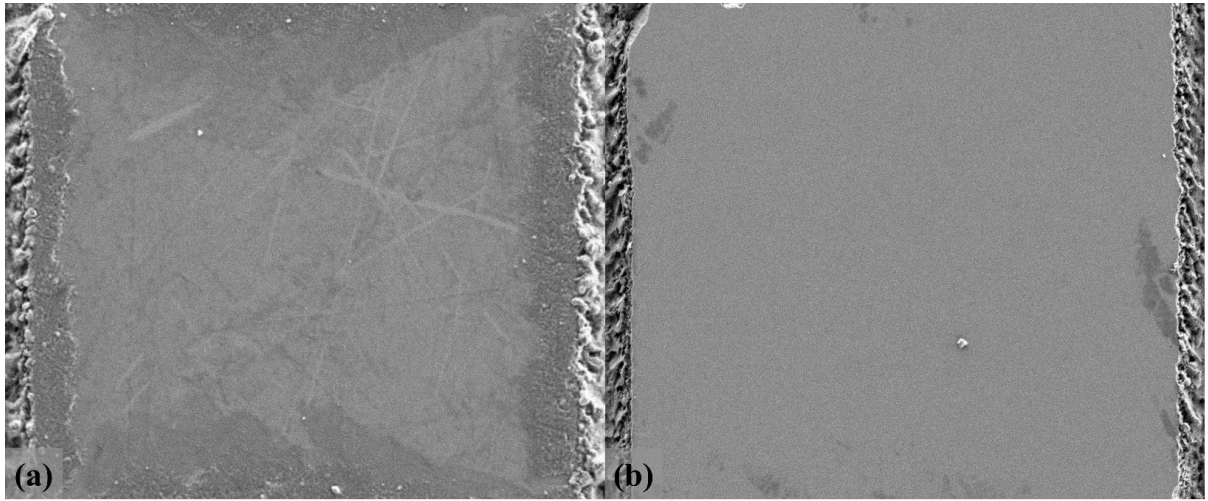
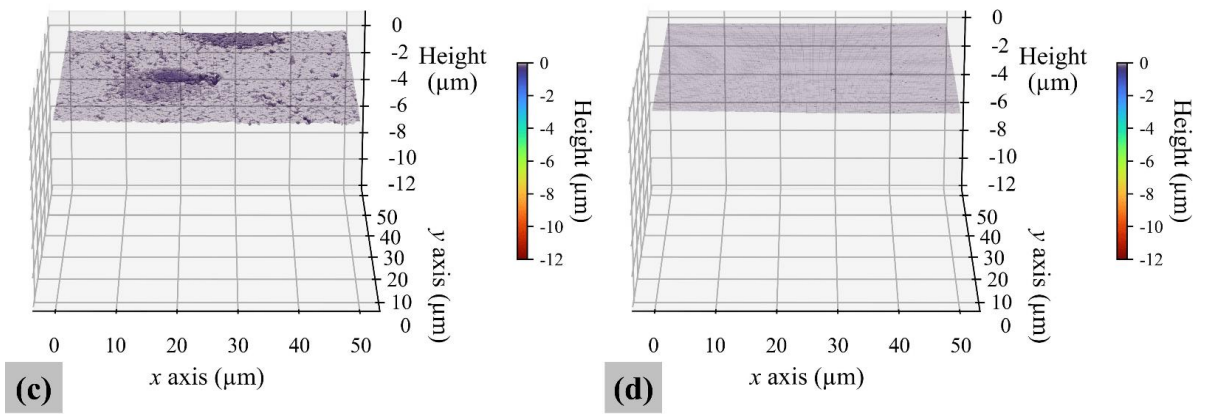
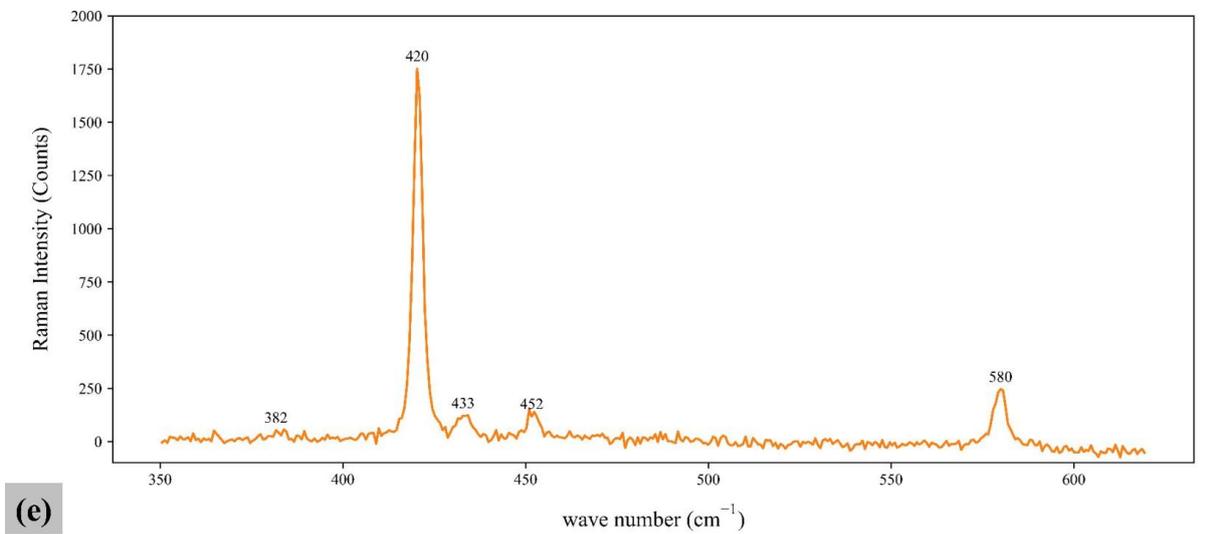

**Fig. S5: Metrology of 320 TW/cm$^2$ and 20 pulses sample.** The measurements of the 320 TW/cm$^2$ and 20 pulses spot include (a) pre-etch and (b) post-etch SEM images, (c) pre-etch and (d) post-etch laser confocal height maps, and (e) pre-etch micro-Raman spectra for the center of the irradiated region.

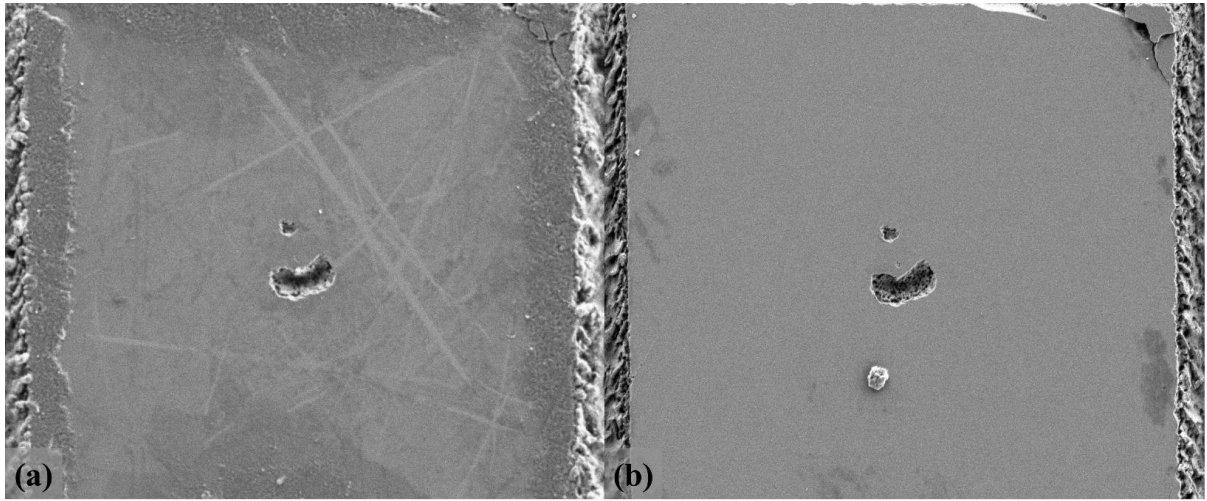
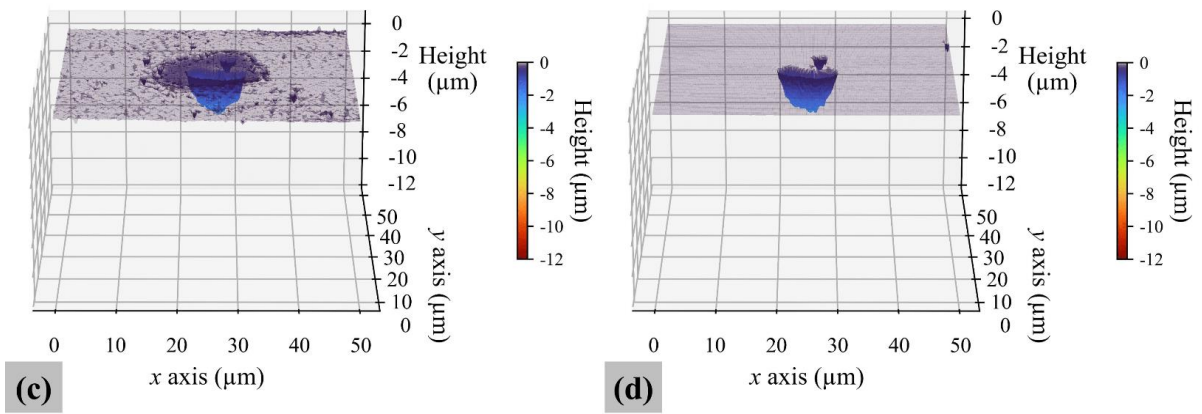
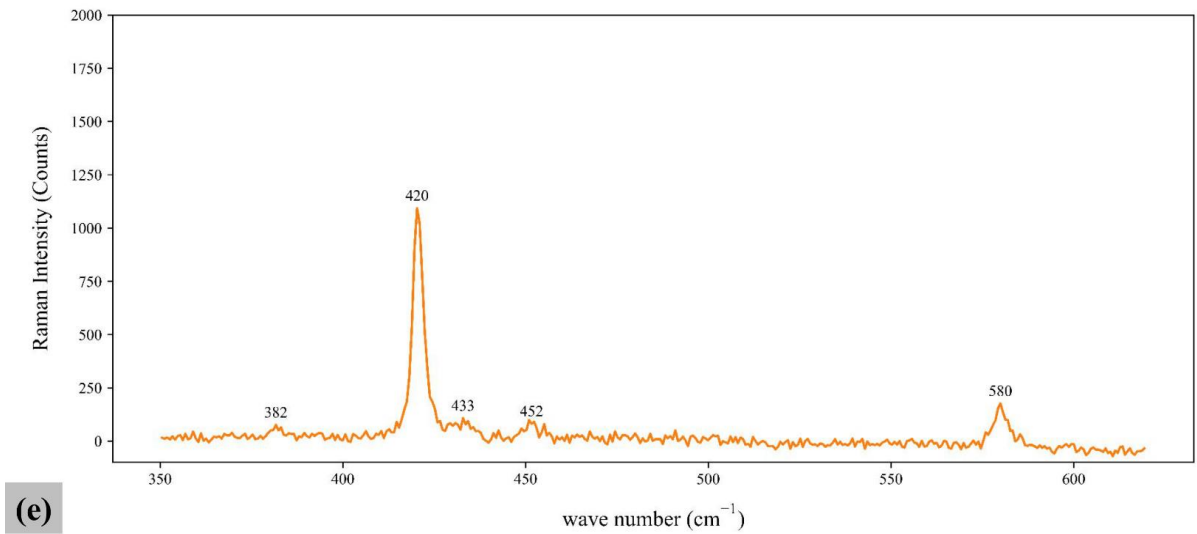

**Fig. S6: Metrology of 640 TW/cm$^2$ and 20 pulses sample.** The measurements of the 640 TW/cm$^2$ and 20 pulses spot include (a) pre-etch and (b) post-etch SEM images, (c) pre-etch and (d) post-etch laser confocal height maps, and (e) pre-etch micro-Raman spectra for the center of the irradiated region.

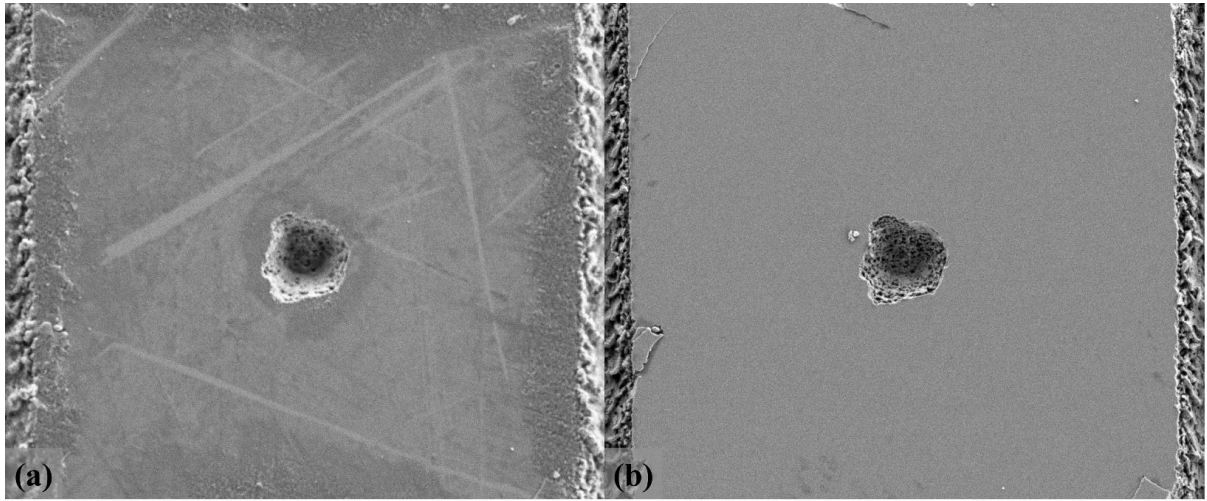
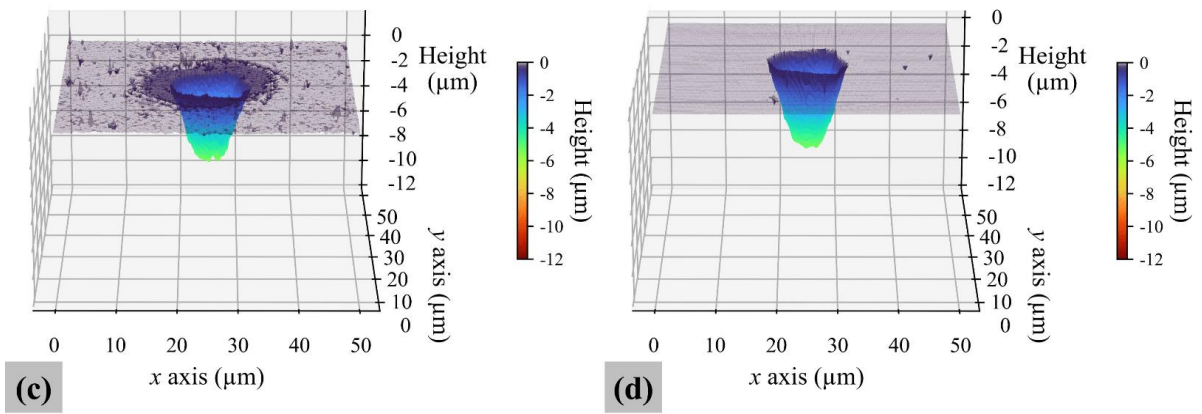
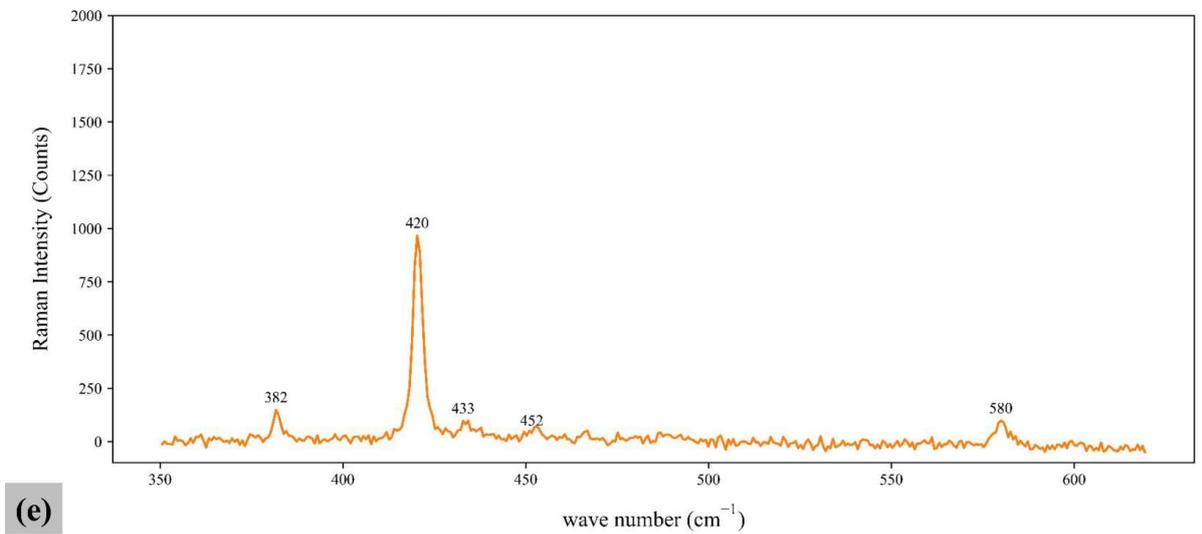

**Fig. S7: Metrology of 1280 TW/cm² and 20 pulses sample.** The measurements of the 1280 TW/cm² and 20 pulses spot include (a) pre-etch and (b) post-etch SEM images, (c) pre-etch and (d) post-etch laser confocal height maps, and (e) pre-etch micro-Raman spectra for the center of the irradiated region.

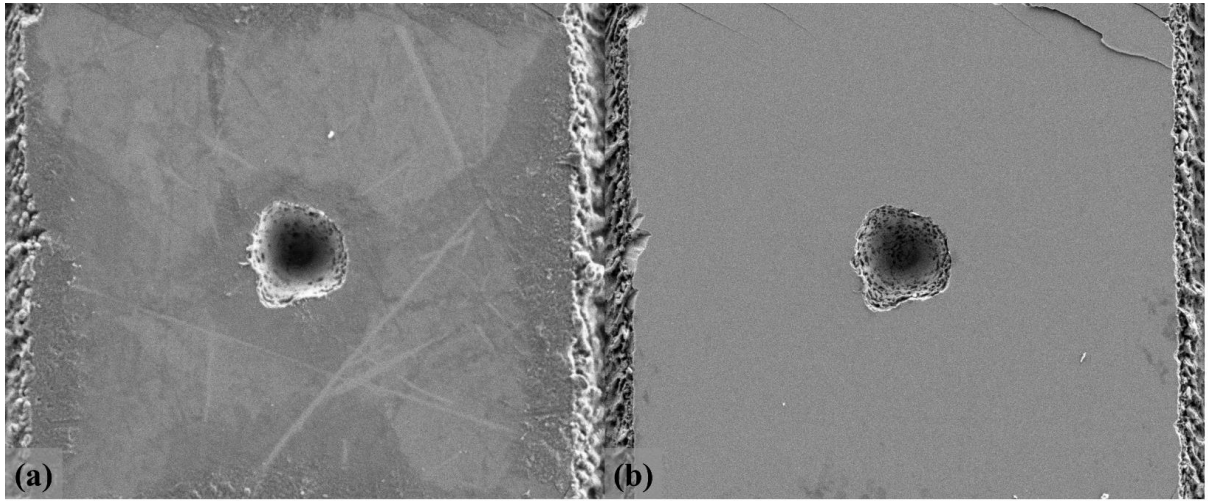
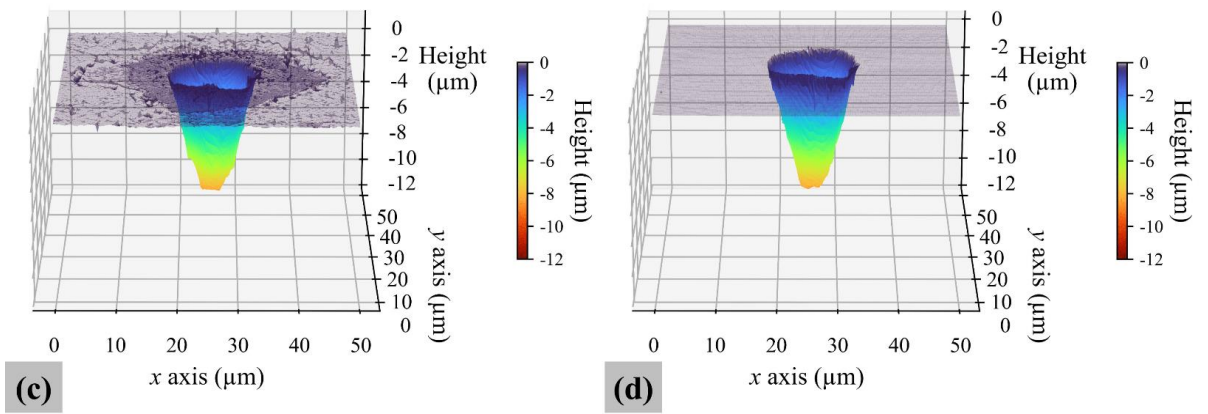
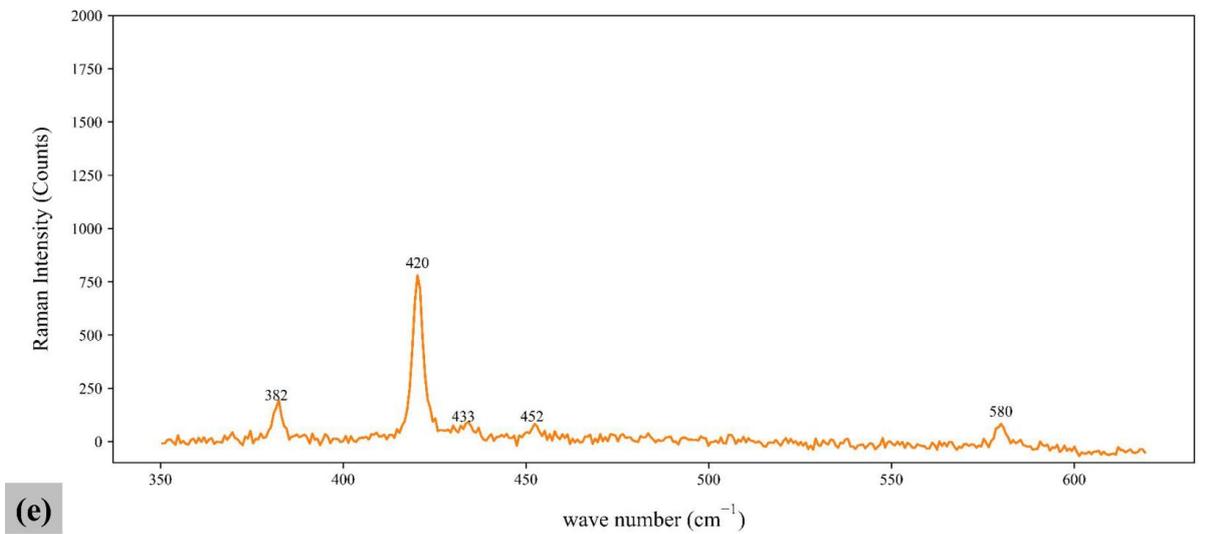

**Fig. S8: Metrology of 2560 TW/cm$^2$ and 20 pulses sample.** The measurements of the 2560 TW/cm$^2$ and 20 pulses spot include (a) pre-etch and (b) post-etch SEM images, (c) pre-etch and (d) post-etch laser confocal height maps, and (e) pre-etch micro-Raman spectra for the center of the irradiated region.

# Supporting Information E: Laser Confocal Measurement of Hierarchical Nanostructures

The post-etch patterned sample of hierarchical nanostructures is analyzed using laser confocal measurements, as displayed in the height map in Fig. S9. Here it can be observed that the fabricated microstructures have a period of 7 μm and height of up to 5.7 μm. The microstructure roughness is calculated from the confocal data and is around 2.34.

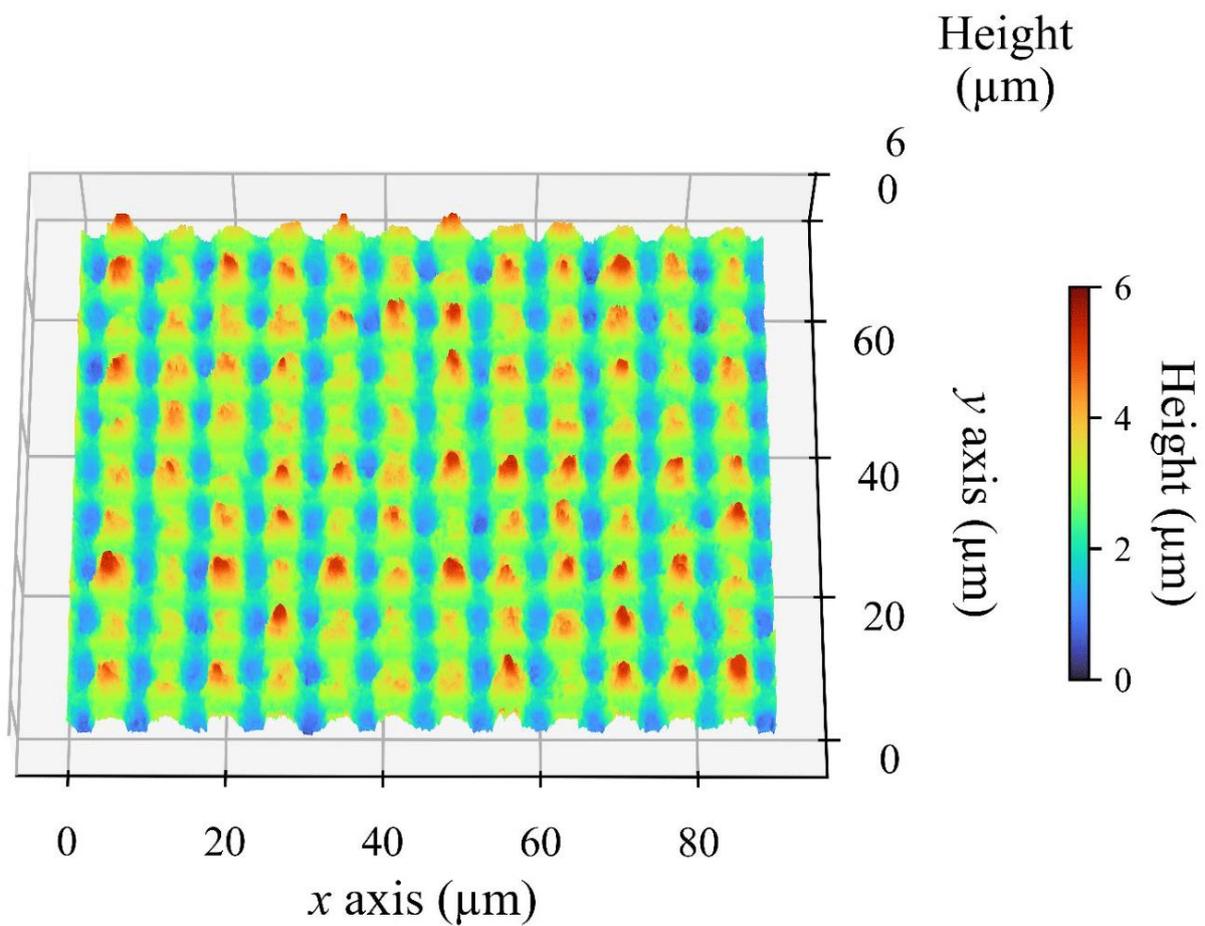

**Fig. S9: Height map of post-etch hierarchical nanostructures.** Created using data from laser confocal measurements.

# Supporting Information F: Static Contact Angle Measurements

Static water contact angle measurements of the silane-coated sapphire substrate and nanostructures are displayed in Fig. S10 below. The substrate in Fig. S10(a) has a measured static contact angle of 106 degrees, while a static contact angle of 140 degrees can be observed from the sapphire nanostructures in Fig. S10(b).

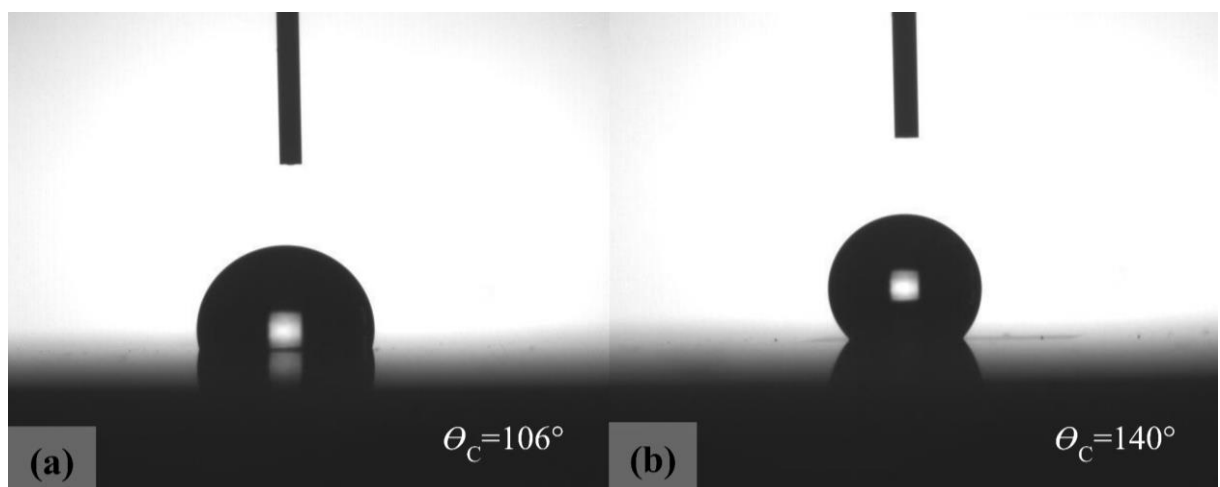

**Fig. S10: Static contact angle measurements.** Images of distilled water droplets deposited on the silane-coated (a) substrate and (b) nanostructures in the goniometer.

# Supporting Information G: Total Transmittance Measurements

The total transmittance of the post-etch substrate and hierarchical nanostructures are displayed in Fig. S11. The substrate displays little variation in transmittance across the measured wavelengths from 250 to 2500 nm, while the nanostructures demonstrate a slight increase in transmittance with increasing wavelength. The total transmission of the nanostructures exceeds that of the substrate at wavelengths above 1601 nm.

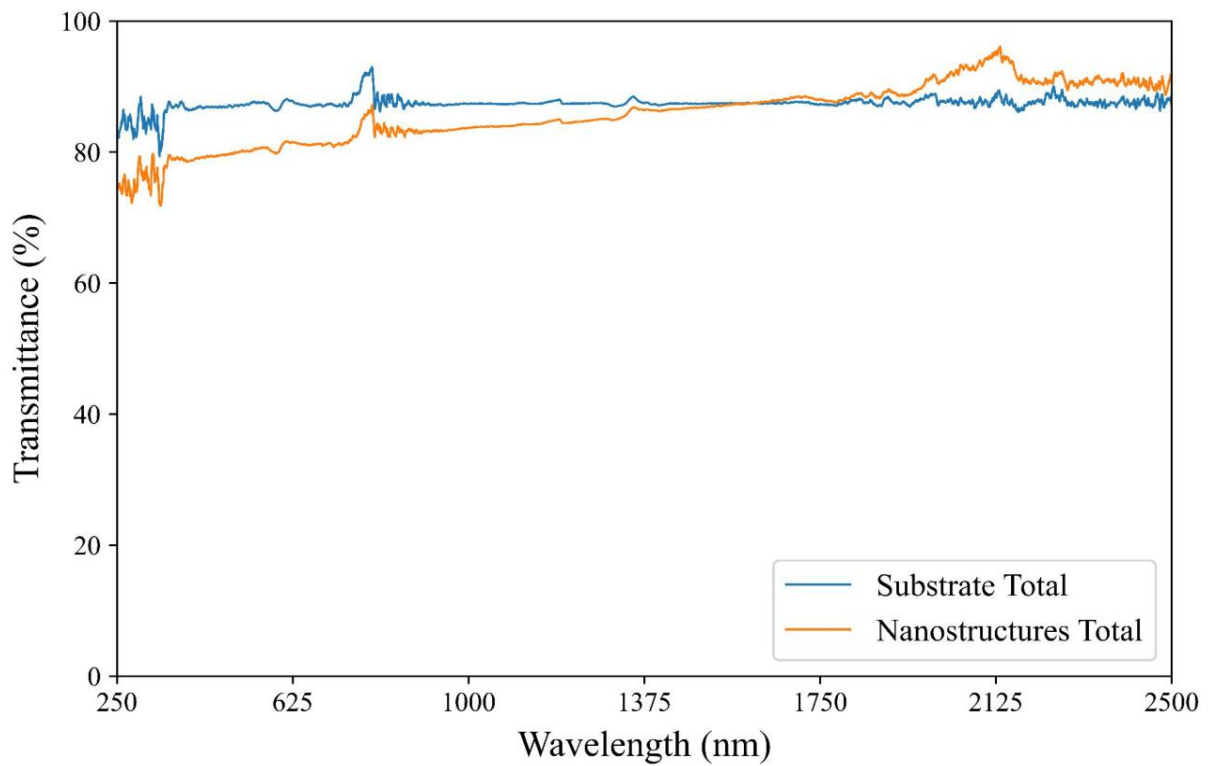

**Fig. S11: Total Transmittance Measurements.** Total transmission measurements of the post-etch sapphire substrate and nanostructures across the UV-Vis-NIR spectrum.